\documentclass[10pt,epsf,preprint]{aastex}
\usepackage{epsf}
\input psfig 

\begin{document}
\title{Early-type galaxies in the SDSS. III. The Fundamental Plane}

\author{
Mariangela Bernardi\altaffilmark{\ref{Chicago},\ref{CMU}},
Ravi K. Sheth\altaffilmark{\ref{Fermilab},\ref{Pitt}},
James Annis\altaffilmark{\ref{Fermilab}},
Scott Burles\altaffilmark{\ref{Fermilab}},
Daniel J. Eisenstein\altaffilmark{\ref{Arizona}},
Douglas P. Finkbeiner\altaffilmark{\ref{Berkeley},\ref{Princeton},\ref{HF}},
David W. Hogg\altaffilmark{\ref{NYU}},
Robert H. Lupton\altaffilmark{\ref{Princeton}},
David J. Schlegel\altaffilmark{\ref{Princeton}}, 
Mark Subbarao\altaffilmark{\ref{Chicago}},
Neta A. Bahcall\altaffilmark{\ref{Princeton}},
John P. Blakeslee\altaffilmark{\ref{JHU}},
J. Brinkmann\altaffilmark{\ref{APO}},
Francisco J. Castander\altaffilmark{\ref{yale},\ref{chile}},
Andrew J. Connolly\altaffilmark{\ref{Pitt}}, 
Istvan Csabai\altaffilmark{\ref{Eotvos},\ref{JHU}},
Mamoru Doi\altaffilmark{\ref{Tokyo1},\ref{Tokyo2}},
Masataka Fukugita\altaffilmark{\ref{ICRR},\ref{IAS}},
Joshua Frieman\altaffilmark{\ref{Chicago},\ref{Fermilab}},
Timothy Heckman\altaffilmark{\ref{JHU}},
Gregory S. Hennessy\altaffilmark{\ref{USNO}},
\v{Z}eljko Ivezi\'{c}\altaffilmark{\ref{Princeton}},
G. R. Knapp\altaffilmark{\ref{Princeton}},
Don Q. Lamb\altaffilmark{\ref{Chicago}},
Timothy McKay\altaffilmark{\ref{UMich}},
Jeffrey A. Munn\altaffilmark{\ref{USNO}},
Robert Nichol\altaffilmark{\ref{CMU}},
Sadanori Okamura\altaffilmark{\ref{Tokyo3},\ref{Tokyo2}}, 
Donald P. Schneider\altaffilmark{\ref{PSU}},
Aniruddha R. Thakar\altaffilmark{\ref{JHU}},
and Donald G.\ York\altaffilmark{\ref{Chicago}}
}

\newcounter{address}
\setcounter{address}{1}
\altaffiltext{\theaddress}{\stepcounter{address}
University of Chicago, Astronomy \& Astrophysics
Center, 5640 S. Ellis Ave., Chicago, IL 60637\label{Chicago}}
\altaffiltext{\theaddress}{\stepcounter{address}
Department of Physics, Carnegie Mellon University, Pittsburgh, PA 15213
\label{CMU}}
\altaffiltext{\theaddress}{\stepcounter{address}
Fermi National Accelerator Laboratory, P.O. Box 500,
Batavia, IL 60510\label{Fermilab}}
\altaffiltext{\theaddress}{\stepcounter{address}
Department of Physics and Astronomy, University of Pittsburgh, Pittsburgh, PA 15620\label{Pitt}}
\altaffiltext{\theaddress}{\stepcounter{address}
Stewart Observatory, University of Arizona, 933 N. Clarry Ave., Tucson, AZ 85121\label{Arizona}}
\altaffiltext{\theaddress}{\stepcounter{address}
Department of Astronomy, University of California at Berkeley, 601 Campbell Hall, Berkeley, CA 94720\label{Berkeley}}
\altaffiltext{\theaddress}{\stepcounter{address}
Princeton University Observatory, Princeton, NJ 08544\label{Princeton}}
\altaffiltext{\theaddress}{\stepcounter{address}Hubble Fellow\label{HF}}
\altaffiltext{\theaddress}{\stepcounter{address}
Department of Physics, New York University, 4 Washington Place, New York, NY 10003\label{NYU}}
\altaffiltext{\theaddress}{\stepcounter{address}
Department of Physics \& Astronomy, The Johns Hopkins University, 3400 North Charles Street, Baltimore, MD 21218-2686\label{JHU}}
\altaffiltext{\theaddress}{\stepcounter{address}
Apache Point Observatory, 2001 Apache Point Road, P.O. Box 59, Sunspot, NM
88349-0059\label{APO}}
\altaffiltext{\theaddress}{\stepcounter{address} Yale University, P. O. Box
208101, New Haven, CT 06520\label{yale}}
\altaffiltext{\theaddress}{\stepcounter{address} Universidad de Chile, Casilla
36-D, Santiago, Chile\label{chile}}
\altaffiltext{\theaddress}{\stepcounter{address}
Department of Physics of Complex Systems, E\"otv\"os University, Budapest, H-1117 Hungary\label{Eotvos}}
\altaffiltext{\theaddress}{\stepcounter{address}
Institute of Astronomy, School of Science, University of Tokyo, Mitaka, Tokyo 181-0015, Japan\label{Tokyo1}}
\altaffiltext{\theaddress}{\stepcounter{address}
Research Center for the Early Universe, School of Science,
    University of Tokyo, Tokyo 113-0033, Japan\label{Tokyo2}}
\altaffiltext{\theaddress}{\stepcounter{address}
Institute for Cosmic Ray Research, University of Tokyo, Kashiwa 277-8582, Japan\label{ICRR}}
\altaffiltext{\theaddress}{\stepcounter{address}
Institute for Advanced Study, Olden Lane, Princeton, NJ 08540\label{IAS}}
\altaffiltext{\theaddress}{\stepcounter{address}
U.S. Naval Observatory, 3450 Massachusetts Ave., NW, Washington, DC 20392-5420\label{USNO}}
\altaffiltext{\theaddress}{\stepcounter{address}
Department of Physics, University of Michigan, 500 East University, Ann Arbor, MI 48109\label{UMich}}
\altaffiltext{\theaddress}{\stepcounter{address}
Department of Astronomy, University of Tokyo,
   Tokyo 113-0033, Japan\label{Tokyo3}}
\altaffiltext{\theaddress}{\stepcounter{address}
Department of Astronomy and Astrophysics, The Pennsylvania State University, University Park, PA 16802\label{PSU}}



\begin{abstract}
A magnitude limited sample of nearly 9000 early-type galaxies, 
in the redshift range $0.01 \le z \le 0.3$, was selected from the 
Sloan Digital Sky Survey  using morphological and spectral criteria.  
The Fundamental Plane relation in this sample is 
 $R_o \propto \sigma^{1.49\pm 0.05}\,I_o^{-0.75\pm 0.01}$ in the $r^*$ 
band.  It is approximately the same in the $g^*$, $i^*$ and $z^*$ bands.  
Relative to the population at the median redshift in the sample, 
galaxies at lower and higher redshifts have evolved only little.  
If the Fundamental Plane is used to quantify this evolution then 
the apparent magnitude limit can masquerade as evolution; once this 
selection effect has been accounted for, the evolution is consistent 
with that of a passively evolving population which formed the bulk of 
its stars about 9~Gyrs ago.  
One of the principal advangtages of the SDSS sample over previous samples 
is that the galaxies in it lie in environments ranging from isolation in 
the field to the dense cores of clusters.  The Fundamental Plane shows 
that galaxies in dense regions are slightly different from galaxies in 
less dense regions.

\end{abstract}  

\section{Introduction}
This is the third of four papers in which the properties of 
$\sim 9000$ early-type galaxies, in the redshift range 
$0.01\le z\le 0.3$ are studied.  
Paper I (Bernardi et al. 2003a) describes how the sample was selected 
from the SDSS database.  The sample is essentially magnitude limited, 
and the galaxies in it span a wide range of environments.  
Each galaxy in the sample has measured values of luminosity $L$, 
size $R_o$ and surface brightness $I_o=(L/2)/R_o^2$ in four bands 
($g^*$, $r^*$, $i^*$ and $z^*$), a velocity dispersion $\sigma$, 
a redshift, and an estimate of the local density.  
Paper~II (Bernardi et al. 2003b) shows that the joint distribution 
of early-type galaxy luminosities, radii and velocity dispersions is 
reasonably well fit by a tri-variate Gaussian.  
It also shows various correlations 
between pairs of variables, such as the luminosity--velocity dispersion 
relation, the luminosity--size relation, and the relation between radius 
and surface brightness.  
Paper~IV (Bernardi et al. 2003c) uses the spectra of these galaxies to 
provide information on the chemical evolution of the early-type population.  

This paper places special emphasis on the Fundamental Plane (FP) 
relation between size, surface brightness and velocity dispersion.  
It shows how the FP depends on waveband, color, redshift and 
environment.  
In Section~\ref{fits} we compare the results of a maximum likelihood 
analysis which can account for evolution and selection effects, 
as well as correlations between errors (e.g., Saglia et al. 2001, 
and Paper~II of this series) 
and standard regression estimates, which cannot.  
Section~\ref{fpscat} checks if the residuals from the plane correlate 
with any other observables.  A discussion of the mass-to-light 
ratio is provided in Section~\ref{m2l}.  Evidence for weak evolution is 
presented in Section~\ref{evolve}, and weak trends with environment 
are found in Section~\ref{environ}.  The distribution of the galaxies 
in our sample in $\kappa$-space (Bender, Burstein \& Faber 1992) 
is shown in Section~\ref{kpspace}.  We summarize our findings in 
Section~\ref{discuss}.

Except where stated otherwise, we write the Hubble constant as 
$H_0=100\,h~\mathrm{km\,s^{-1}\,Mpc^{-1}}$, and we perform our 
analysis in a cosmological world model with 
$(\Omega_{\rm M},\Omega_{\Lambda},h)=(0.3,0.7,0.7)$, where 
$\Omega_{\rm M}$ and $\Omega_{\Lambda}$ are the present-day scaled 
densities of matter and cosmological constant.  
In such a model, the age of the Universe at the present time is 
$t_0=9.43h^{-1}$ Gyr.  For comparison, an Einstein-de Sitter model has 
$(\Omega_{\rm M},\Omega_{\Lambda})=(1,0)$ and $t_0=6.52h^{-1}$~Gyr.  
We frequently use the notation $h_{70}$ as a reminder that we have 
set $h=0.7$.  Also, we will frequently be interested in the 
logarithms of physical quantities.  Our convention is to set 
$R\equiv\log_{10}R_o$ and $V\equiv \log_{10}\sigma$, where $R_o$ 
and $\sigma$ are effective radii in $~h_{70}^{-1}$~kpc and velocity 
dispersions in km~s$^{-1}$, respectively.

\section{The Fundamental Plane}\label{fp}
In any given band, each galaxy in our sample is characterized by 
three numbers:  its luminosity, $L$, its size, $R_o$, and its velocity 
dispersion, $\sigma$.  
Correlations between these three observables are expected if 
early-type galaxies are in virial equilibrium, because 
\begin{equation}
\sigma_{vir}^2 \propto {GM_{vir}\over 2R_{vir}} 
                   \propto \left({M_{vir}\over L}\right)R_{vir}
\left({L/2\over R_{vir}^2}\right).
\end{equation}
If the size parameter $R_{vir}$ which enters the virial theorem is 
linearly proportional to the observed effective radius of the light, 
$R_o$, and if the observed line-of-sight velocity dispersion $\sigma$ 
is linearly proportional to $\sigma_{vir}$, then this relates the 
observed velocity dispersion to the product of the observed surface 
brightness and effective radius.  
Following Djorgovski \& Davis (1987), correlations involving all 
three variables are often called the Fundamental Plane (FP).  
In what follows, we will show how the surface brightness, $R_o$, and 
$\sigma$ are correlated.  Because both 
$\mu_o\propto -2.5\log_{10}[(L/2)/R_o^2]$ and $\sigma$ are distance 
independent quantities (this assumes that cosmological dimming and 
K-corrections have been computed correctly), it is in these variables 
that studies of early-type galaxies are usually presented.  

\subsection{Finding the best-fitting plane}\label{fits}
The Fundamental Plane is defined by:
\begin{equation}
\log_{10} R_o = a\,\log_{10}\sigma + b\,\log_{10}I_o + c
\end{equation}
where the coefficients $a$, $b$, and $c$ are determined by minimizing 
the residuals from the plane.  There are a number of ways in which 
this is usually done.  Let 
\begin{eqnarray}
\Delta_1 &\equiv& 
 \log_{10}R_o - a\,\log_{10}\sigma - b\,\log_{10}I_o - c
 \qquad {\rm and}\nonumber \\ \nonumber \\
\Delta_o &\equiv& {\Delta_1\over (1 + a^2 + b^2)^{1/2}}.  
\label{deltas}
\end{eqnarray}
Then summing $\Delta_1^2$ over all $N$ galaxies and finding that set of 
$a$, $b$ and $c$ for which the sum is minimized gives what is often 
called the direct fit, whereas minimizing the sum of $\Delta_o^2$ 
instead gives the orthogonal fit.  Although the orthogonal fit is, 
perhaps, the more physically meaningful, the direct fit is of more 
interest if the FP is to be used as a distance indicator.  

A little algebra shows that the direct fit coefficients are 
\begin{eqnarray}
a &=& {\sigma_{II}^2\sigma_{RV}^2-\sigma_{IR}^2\sigma_{IV}^2\over
\sigma_{II}^2\sigma_{VV}^2-\sigma_{IV}^4}, \qquad 
b = {\sigma_{VV}^2\sigma_{IR}^2-\sigma_{RV}^2\sigma_{IV}^2\over
\sigma_{II}^2\sigma_{VV}^2-\sigma_{IV}^4}, \nonumber\\ 
c &=& \overline{\log_{10}R_o} - a\,\overline{\log_{10}\sigma} - 
b\,\overline{\log_{10}I_o}, \qquad {\rm and} \nonumber \\
\langle \Delta_1^2\rangle &=& 
{\sigma_{II}^2\sigma_{RR}^2\sigma_{VV}^2 
- \sigma_{II}^2 \sigma_{RV}^4 - \sigma_{RR}^2 \sigma_{IV}^4 - 
\sigma_{VV}^2 \sigma_{IR}^4 + 2 \sigma_{IR}^2\sigma_{IV}^2\sigma_{RV}^2
\over \sigma_{II}^2\sigma_{VV}^2-\sigma_{IV}^4},
\label{dfit}
\end{eqnarray}
where $\overline{\log_{10}X}\equiv \sum_i \log_{10}X_i/N$ and 
$\sigma^2_{xy}\equiv \sum_i (\log_{10}X_i - \overline{\log_{10}X})
(\log_{10}Y_i - \overline{\log_{10}Y})/N$, and 
$X$ and $Y$ can be $I_o$, $R_o$ or $\sigma$.  
For what follows, it is also convenient to define 
 $r_{xy} = \sigma_{xy}^2/(\sigma_{xx}\sigma_{yy})$.  
The final expression above gives the scatter around the relation.  
If surface brightness and velocity dispersion are uncorrelated 
(we will show below that, indeed, $\sigma_{IV}\approx 0$), then $a$ 
equals the slope of the relation between velocity dispersion and the 
mean size at fixed velocity dispersion, $b$ is the slope of the relation 
between surface-brightness and the mean size at fixed surface-brightness, 
and the rms scatter is $\sigma_{RR}\sqrt{1 - r_{RV}^2 - r_{IR}^2}$.  
Errors in the observables affect the measured $\sigma_{xy}^2$, and thus 
will bias the determination of the best-fit coefficients and the intrinsic 
scatter around the fit.  If $\epsilon_{xy}$ is the rms error in the joint 
measurement of $\log_{10} X$ and $\log_{10}Y$, then subtracting the 
appropriate $\epsilon^2_{xy}$ from each $\sigma_{xy}^2$ before using them 
provides estimates of the error-corrected values of $a$, $b$ and $c$.  
Expressions for the orthogonal fit coefficients can be derived similarly, 
although, because they require solving a cubic equation, they are lengthy, 
so we have not included them here.  

Neither minimization procedure above accounts for the fact that the 
sample is magnitude-limited, and has a cut at small velocity dispersions.  
In addition, because our sample spans a wide range of redshifts, we must 
worry about effects which may be due to evolution.  The magnitude limit 
means that we cannot simply divide our sample up into small redshift 
ranges (over which evolution is negligible), because a small redshift 
range probes only a limited range of luminosities, sizes and velocity 
dispersions.  To account for all these effects, we use the 
maximum-likelihood approach (e.g. Saglia et al. 2001) described in 
Paper~II.  
This method is the natural choice given that the joint distribution of 
$M\equiv -2.5\log_{10}L$, $R\equiv\log_{10}R_o$ and $V\equiv\log_{10}\sigma$ 
is quite well described by a multivariate Gaussian.  The maximum 
likelihood estimates of the mean values of these variables, and the 
parameters of the covariance matrix ${\cal C}$ which describes the 
correlations between these variables are shown in Table~1 of Paper~II.  
What remains is to write down how the covariance matrix changes when 
we change variables from $(M,R,V)$ to $(\mu,R,V)$.  Because 
$(\mu_o-\mu_*) \equiv (M-M_*) + 5(R-R_*)$, the covariance matrix 
becomes 
\begin{eqnarray}
{\cal F} &\equiv& \left( \begin{array}{ccc} 
     \sigma^2_M + 10\sigma_M\sigma_R\rho_{RM} + 25\sigma_R^2 & 
     \sigma_R\sigma_M\,\rho_{RM} + 5\sigma^2_R & 
     \sigma_V\sigma_M\,\rho_{VM} + 5\sigma_R\sigma_V\,\rho_{RV}\\
     \sigma_R\sigma_M\,\rho_{RM} + 5\sigma^2_R & 
     \sigma^2_R & \sigma_R\sigma_V\,\rho_{RV}\\
     \sigma_V\sigma_M\,\rho_{VM} + 5\sigma_R\sigma_V\,\rho_{RV} & 
     \sigma_R\sigma_V\,\rho_{RV} & \sigma^2_V\\
           \end{array}\right) \nonumber \\ 
   &\equiv& \left( \begin{array}{ccc} 
     \sigma_\mu^2 & 
     \sigma_R\sigma_\mu\,\rho_{R\mu} & 
     \sigma_V\sigma_\mu\,\rho_{V\mu}\\
     \sigma_R\sigma_\mu\,\rho_{R\mu} & 
     \sigma^2_R & \sigma_R\sigma_V\,\rho_{RV}\\
     \sigma_V\sigma_\mu\,\rho_{V\mu} & 
     \sigma_R\sigma_V\,\rho_{RV} & \sigma^2_V\\
           \end{array}\right) .
\end{eqnarray}
The coefficients of ${\cal F}$ are given in Table~\ref{MLcov}; 
they were obtained by inserting the values shown in Table~1 
of Paper~II into the first of the equalities above.  
Note that $\mu = -2.5\log_{10}I_o$.  

\begin{table}[t]
\centering
\caption[]{Maximum-likelihood estimates, in the four SDSS bands, of the 
joint distribution of luminosities, sizes and velocity dispersions.  
The mean values of the variables at redshift $z$ are $\mu_*-Qz$, $R_*$, 
$V_*$, and the elements of the covariance matrix ${\cal F}$ defined by 
the various pairwise correlations between the variables are shown.  
These coefficients were obtained from the coefficients of the 
covariance matrix ${\cal C}$ shown in Table~1 in Paper~II.  \\}
\begin{tabular}{cccccccccccc}
\tableline 
Band & $N_{\rm gals}$ & $\mu_*$ & $\sigma_\mu$ & $R_*$ & $\sigma_R$ &
$V_*$ & $\sigma_V$ & $\rho_{R\mu}$ & $\rho_{V\mu}$ & $\rho_{RV}$ & Q\\
\hline\\
$g^*$ & 5825 & $20.74$ & 0.654 & 0.520 & 0.254 & 2.197 & 0.113 & $0.801$ &
$0.005$ & 0.536 & 1.15 \\
$r^*$ & 8228 & $19.87$ & 0.610 & 0.490 & 0.241 & 2.200 & 0.111 & $0.760$ &
$0.000$ & 0.543 & 0.85 \\
$i^*$ & 8022 & $19.40$ & 0.600 & 0.465 & 0.241 & 2.201 & 0.110 & $0.753$ &
$-0.001$ & 0.542 & 0.75 \\
$z^*$ & 7914 & $18.99$ & 0.604 & 0.450 & 0.241 & 2.200 & 0.110 & $0.759$ &
$-0.001$ & 0.543 & 0.60 \\
\tableline
\end{tabular}
\label{MLcov}
\end{table}

This matrix is fundamentally useful because it describes the intrinsic 
correlations between the sizes, surface-brightnesses and velocity 
dispersions of early-type galaxies---the effects of how the sample was 
selected and observational errors have been accounted for.  For example, 
the coefficients in the top right and bottom left of ${\cal F}$ are 
very close to zero, indicating that surface brightness and velocity 
dispersion are uncorrelated.  
In addition, the eigenvalues and vectors of ${\cal F}$ give 
information about the shape and thickness of the Fundamental Plane.  
For example, in $r^*$, the eigenvalues are 0.639, 0.179, and 0.052;  
the smallest eigenvalue is considerably smaller than the other two 
indicating that, when viewed in the appropriate projection, the plane 
is quite thin.  The associated eigenvector gives the coefficients of 
the `orthogonal' fit, and the rms scatter around this orthogonal fit 
is given by the (square root of the) smallest eigenvalue 
(e.g. Saglia et al. 2001).  

If we wish to use the FP as a distance indicator, then we are more 
interested in finding those coefficients which minimize the scatter in 
$R_o$.  This means that we would like to find that pair $(a,b)$ which 
minimize $\langle\Delta_1^2\rangle$, where $\Delta_1$ is given by 
equation~(\ref{deltas}).  A little algebra shows that the solution is 
given by inserting the maximum likelihood estimates of the scatter in 
surface-brightnesses, sizes and velocity dispersions into 
equation~(\ref{dfit}).  

\begin{table}
\centering
\caption[]{Coefficients of the FP in the four SDSS bands.  For each 
set of coefficients, the scatter orthogonal to the plane and in the 
direction of $R_o$ are also given.  
\\}
\begin{tabular}{cccccc}
\tableline 
Band & $a$ & $b$ & $c$ & rms$_{\rm orth}^{\rm int}$ & rms$_{R_o}^{\rm int}$ \\
\hline\\
\\[-8mm]
\multicolumn{6}{c}{\bf Orthogonal fits} \\
\multicolumn{6}{l}{\bf Maximum Likelihood} \\
\\[-6mm]
$g^*$ & 1.45$\pm 0.06$  & $-0.74\pm 0.01$  & $-8.779\pm 0.029$ & 0.056 & 0.100 \\
$r^*$ & 1.49$\pm 0.05$  & $-0.75\pm 0.01$  & $-8.778\pm 0.020$ & 0.052 & 0.094 \\
$i^*$ & 1.52$\pm 0.05$  & $-0.78\pm 0.01$  & $-8.895\pm 0.021$ & 0.049 & 0.091 \\
$z^*$ & 1.51$\pm 0.05$  & $-0.77\pm 0.01$  & $-8.707\pm 0.023$ & 0.049 & 0.089 \\
\multicolumn{6}{l}{\bf $\chi^2 - $ Evolution $-$ Selection effects}\\
\\[-5mm]
$g^*$ & 1.43$\pm 0.06$  & $-0.78\pm 0.01$  & $-9.057\pm 0.032$ & 0.058 & 0.101 \\
$r^*$ & 1.45$\pm 0.05$  & $-0.76\pm 0.01$  & $-8.719\pm 0.020$ & 0.052 & 0.094 \\
$i^*$ & 1.48$\pm 0.05$  & $-0.77\pm 0.01$  & $-8.699\pm 0.024$ & 0.050 & 0.090 \\
$z^*$ & 1.48$\pm 0.05$  & $-0.77\pm 0.01$  & $-8.577\pm 0.025$ & 0.049 & 0.089\\
\multicolumn{6}{l}{\bf $\chi^2 - $ Evolution}\\
\\[-5mm]
$g^*$ & 1.35$\pm 0.06$  & $-0.77\pm 0.01$  & $-8.820\pm 0.033$ & 0.058 & 0.100 \\
$r^*$ & 1.40$\pm 0.05$  & $-0.77\pm 0.01$  & $-8.678\pm 0.023$ & 0.053 & 0.092 \\
$i^*$ & 1.41$\pm 0.05$  & $-0.78\pm 0.01$  & $-8.688\pm 0.024$ & 0.050 & 0.090 \\
$z^*$ & 1.41$\pm 0.05$  & $-0.78\pm 0.01$  & $-8.566\pm 0.026$ & 0.048 & 0.089\\
\hline\\
\\[-8mm]
\multicolumn{6}{c}{\bf Direct fits} \\
\multicolumn{6}{l}{\bf Maximum Likelihood} \\
\\[-6mm]
$g^*$ & 1.08$\pm 0.05$ & $-0.74\pm 0.01$ & $-8.033\pm 0.024$ & 0.061 & 0.092\\
$r^*$ & 1.17$\pm 0.04$ & $-0.75\pm 0.01$ & $-8.022\pm 0.020$ & 0.056 & 0.088\\
$i^*$ & 1.21$\pm 0.04$ & $-0.77\pm 0.01$ & $-8.164\pm 0.018$ & 0.053 & 0.085\\
$z^*$ & 1.20$\pm 0.04$ & $-0.76\pm 0.01$ & $-7.995\pm 0.021$ & 0.053 & 0.084\\
\multicolumn{6}{l}{\bf $\chi^2 - $ Evolution $-$ Selection effects}\\
\\[-5mm]
$g^*$ & 1.05$\pm 0.05$ & $-0.79\pm 0.01$ & $-8.268\pm 0.026$ & 0.063 & 0.094\\
$r^*$ & 1.12$\pm 0.04$ & $-0.76\pm 0.01$ & $-7.932\pm 0.020$ & 0.057 & 0.088\\
$i^*$ & 1.14$\pm 0.04$ & $-0.76\pm 0.01$ & $-7.904\pm 0.019$ & 0.054 & 0.085\\
$z^*$ & 1.14$\pm 0.04$ & $-0.76\pm 0.01$ & $-7.784\pm 0.021$ & 0.053 & 0.084\\
\multicolumn{6}{l}{\bf $\chi^2 - $ Evolution}\\
\\[-5mm]
$g^*$ & 0.99$\pm 0.05$ & $-0.76\pm 0.01$ & $-7.921\pm 0.026$ & 0.065 & 0.093\\
$r^*$ & 1.06$\pm 0.04$ & $-0.75\pm 0.01$ & $-7.775\pm 0.020$ & 0.059 & 0.088\\
$i^*$ & 1.09$\pm 0.04$ & $-0.77\pm 0.01$ & $-7.823\pm 0.018$ & 0.056 & 0.085\\
$z^*$ & 1.09$\pm 0.04$ & $-0.78\pm 0.01$ & $-7.818\pm 0.020$ & 0.053 & 0.083\\
\tableline
\end{tabular}
\label{fpcoeffs} 
\end{table}

\begin{figure}
\centering
\epsfxsize=\hsize\epsffile{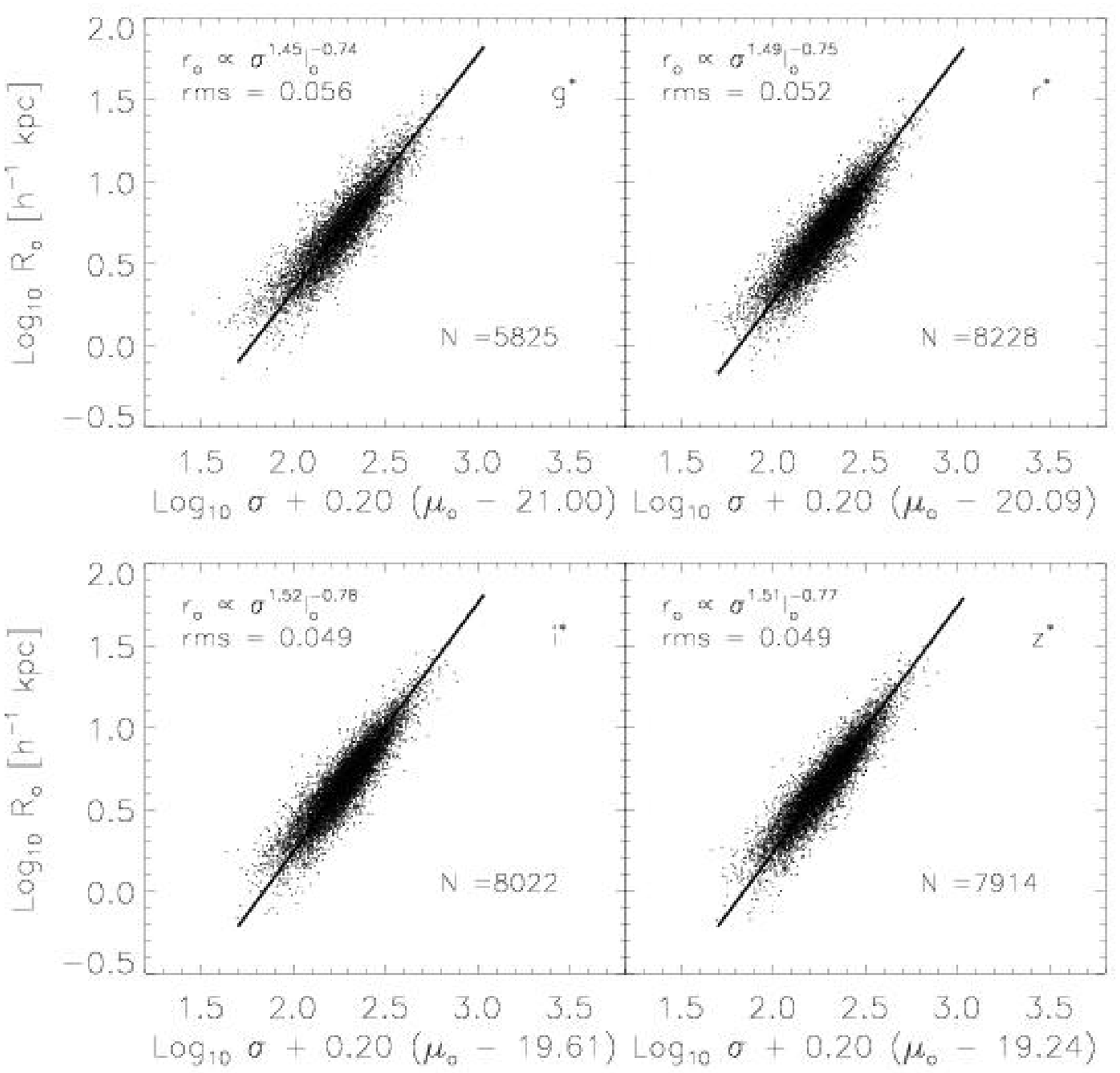}
\caption{The Fundamental Plane in the four SDSS bands.  Coefficients 
shown are those which minimize the scatter orthogonal to the plane, as 
determined by the maximum-likelihood method.  Surface-brightnesses have 
been corrected for evolution.}  
\label{fig:FPgriz}
\end{figure}

The maximum likelihood ${\cal F}$ can be used to provide estimates of 
the direct and orthogonal fit coefficients, as well as the intrinsic 
scatter around the mean relations (orthogonal to the plane as well 
as in the direction of $R_o$).  These are given in Table~\ref{fpcoeffs}.  
Although $b$ is approximately the same both for the `orthogonal' and 
the `direct' fits, $a$ from the direct fit is always about 25\% smaller 
than from the orthogonal fit.  
In either case, note how similar $a$ and $b$ are in all four bands.  
This similarity, and the fact that the thickness of the FP decreases 
slightly with increasing wavelength, can be used to constrain models of 
how different stellar populations (which may contribute more or less to 
the different bands) are distributed in early-type galaxies.  If the direct 
fit is used as a distance indicator, then the thickness of the FP translates 
into an uncertainty in derived distances of about 20\%.  

Table~\ref{fpcoeffs} also shows results from the more traditional 
$\chi^2-$fitting techniques, which were obtained as follows.  (These 
fits were not weighted by errors, and the intrinsic scatter with 
respect to the fits was estimated by subtracting the measurement 
errors in quadrature from the observed scatter.)  
Ignoring evolution and selection effects when minimizing 
$\langle\Delta_1^2\rangle$ and $\langle\Delta_o^2\rangle$, results in 
coefficients $a$ which are about 10\% larger than those we obtained from 
the maximum likelihood method.  We have not shown these in the Table 
for the following reason.  If the population at high redshift is more 
luminous than that nearby, as expected if the evolution is passive, then 
the higher redshift population would have systematically smaller values of 
$\mu_o$.  Since the higher redshift population makes up most of the 
large $R_o$ part of our sample, this could make the Plane appear 
steeper, i.e., it could cause the best-fit $a$ to be biased to a 
larger value.  
If we use the maximum-likelihood estimate of how the luminosities brighten 
with redshift, then we can subtract off the brightening from $\mu_o$ 
before minimizing $\langle\Delta_1^2\rangle$ and $\langle\Delta_o^2\rangle$.  
This reduces the best-fit value of $a$ so that it is closer to that of 
the maximum likelihood method.  The coefficients obtained in this way 
are labeled `$\chi^2$ $-$ Evolution' in Table~\ref{fpcoeffs}; they are 
statistically different from the maximum likelihood estimates, presumably 
because they do not account for selection effects or for the effects of 
observational errors.  
If we weight each galaxy by the inverse of $S(z_i|M_*,Q)$ (the selection 
function defined in Paper~II) when minimizing, then this should at 
least partially account for selection effects.  The resulting estimates 
of $a$, $b$ and $c$ are labeled 
`$\chi^2$ $-$ Evolution $-$ Selection effects' in Table~\ref{fpcoeffs}.   
The small remaining difference between these and the maximum likelihood 
estimates is likely due to the fact that the likelihood analysis 
accounts more consistently for errors.  

Figure~\ref{fig:FPgriz} shows the FP in the four SDSS bands.  
We have chosen to present the plane using the coefficients, 
obtained using the maximum-likelihood method, which minimize the 
scatter orthogonal to the plane.  (In all cases, the evolution of 
the luminosities has been subtracted from the surface brightnesses.)   
The results to follow regarding the shape of the FP, and estimates of how 
the mean properties of early-types depend on redshift and environment, 
are independent of which fits we use.  A fair number of the galaxies in 
our sample have velocity dispersion measurements with small S/N 
(see, e.g., Figure~19 of Paper~I).  The FP is relatively insensitive 
to these objects:  removing objects with S/N $<$ 15 had little effect on 
the best fit values of $a$, $b$.  Removing objects with small axis ratios 
also had little effect on the maximum likelihood coefficients.  

In principle, the likelihood analysis provides an estimate of the error on 
each of the derived coefficients.  However, this estimate assumes that the 
parametric Gaussian form is indeed a good fit.  Although we present 
evidence in Paper~II that the Gaussian form is indeed good, we emphasize 
that, when the data set is larger a non-parametric fit should be 
performed.  Therefore, we have estimated errors on the numbers quoted 
in Table~\ref{fpcoeffs} as follows.  
The large size of our sample allows us to construct many random 
subsamples, each of which is substantially larger than most of the 
samples available in the literature.  Estimating the elements of the 
covariance matrix presented in Table~\ref{MLcov}, and then transforming 
to get the FP coefficients in Table~\ref{fpcoeffs}, in each of these 
subsamples provides an estimate of how well we have determined 
$a$, $b$ and $c$.  (Note that the errors we find in this way are 
comparable to those sometimes quoted in the literature, even though 
each of the subsamples we generated is an order of magnitude larger 
than any sample available in the literature.)  
Because each subsample contains fewer galaxies than our full sample, 
this procedure is likely to provide an overestimate of the true formal 
error for our sample.  However, the formal error does not account for 
the uncertainties in our K-corrections and velocity dispersion aperture 
corrections (discussed in more detail in Paper~I), so an overestimate 
is probably more realistic.  

\begin{table}[t]
\centering
\caption[]{Coefficients of the FP in the complete and magnitude-limited 
simulated catalogs, obtained by minimizing a $\chi^2$ in which evolution 
in the surface brightnesses has been removed, and which weights objects 
by the inverse of the selection function.\\}
\begin{tabular}{cccccc}
\tableline 
Band & $a$ & $b$ & $c$ & rms$_{\rm orth}$ & rms$_{R_o}$ \\
\hline\\
\\[-8mm]
\multicolumn{6}{c}{\bf Orthogonal fits} \\
\multicolumn{6}{l}{\bf Complete} \\
\\[-5mm]
$g^*$ & 1.44$\pm 0.05$ & $-0.74\pm 0.01$ & $-8.763\pm 0.028$ & 0.056 & 0.100 \\
$r^*$ & 1.48$\pm 0.05$ & $-0.75\pm 0.01$ & $-8.722\pm 0.020$ & 0.052 & 0.094 \\
\\[-4mm]
\multicolumn{6}{l}{\bf Magnitude limited}\\
\\[-5mm]
$g^*$ & 1.39$\pm 0.06$ & $-0.74\pm 0.01$ & $-8.643\pm 0.028$ & 0.056 & 0.100 \\
$r^*$ & 1.43$\pm 0.05$ & $-0.76\pm 0.01$ & $-8.721\pm 0.021$ & 0.052 & 0.093 \\
\hline\\
\\[-8mm]
\multicolumn{6}{c}{\bf Direct fits} \\
\multicolumn{6}{l}{\bf Complete} \\
\\[-5mm]
$g^*$ & 1.09$\pm 0.04$ & $-0.74\pm 0.01$ & $-7.992\pm 0.023$ & 0.061 & 0.091\\
$r^*$ & 1.16$\pm 0.04$ & $-0.75\pm 0.01$ & $-8.005\pm 0.020$ & 0.056 & 0.088\\
\\[-4mm]
\multicolumn{6}{l}{\bf Magnitude limited}\\
\\[-5mm]
$g^*$ & 1.04$\pm 0.05$ & $-0.74\pm 0.01$ & $-7.817\pm 0.025$ & 0.061 & 0.090 \\
$r^*$ & 1.11$\pm 0.04$ & $-0.75\pm 0.01$ & $-7.895\pm 0.020$ & 0.056 & 0.087 \\
\tableline
\end{tabular}
\label{simcoeffs} 
\end{table}

As a check on the relative roles of evolution and selection effects, 
we simulated complete and magnitude-limited samples (with a velocity 
dispersion cut) following the procedures outlined in Appendix~A of Paper~II.  
We then estimated the coefficients of the FP in the simulated catalogs 
using the different methods.  The results are summarized in 
Table~\ref{simcoeffs}.  When applied to the complete simulations, the 
$\chi^2-$minimization method yields estimates of $a$ which are biased 
high; it yields the input Fundamental Plane coefficients only after 
evolution has been subtracted from the surface brightnesses.  
However, in the magnitude limited simulations, once evolution has been 
subtracted, it provides an estimate of $a$ which is biased low, unless 
selection effects are also accounted for.  Note that this is similar to 
what we found with the data.  The maximum-likelihood method successfully 
recovers the same intrinsic covariance matrix and evolution as the one 
used to generate the simulations, both for the complete and the 
magnitude-limited mock catalogs, and so it recovers the same correct 
coefficients for the FP in both cases.  (We have not shown these estimates 
in the Table.)

\begin{table}[t]
\centering
\caption[]{Selection of Fundamental Plane coefficients from the literature.\\}
\begin{tabular}{lllllll}
\tableline
Source & Band & N$_{\rm gal}$ & $a$ & $b$ & $\Delta_{R_o}$ & Fit method\\
\hline\\
\\[-4mm]
Dressler et al. (1987) & $B$ & 97 & 1.33$\pm 0.05$ & $-0.83\pm 0.03$ & 20\% & inverse\\
Lucey et al. (1991) & $B$ & 26 & 1.27$\pm0.07$ & $-0.78\pm 0.09$ & 13\% & inverse\\
Guzm\'an et al. (1993) & $V$ & 37 & 1.14$\pm$ -- & $-0.79\pm$ -- & 17\% &
direct \\
Kelson et al. (2000) & $V$ & 30 & 1.31$\pm 0.13$ & $-0.86\pm 0.10$ & 14\% &
orthogonal \\
Djorgovski \& Davis (1987) & $r_{G}$ & 106 & 1.39$\pm 0.14$ & $-0.90\pm 0.09$ & 20\% &
2-step inverse \\
J{\o}rgensen et al. (1996) & $r$ & 226 & 1.24$\pm 0.07$ & $-0.82\pm
0.02$ & 19\% & orthogonal \\
Hudson et al. (1997) & $R$ & 352 & 1.38$\pm 0.04$ & $-0.82\pm 0.03$ & 20\% & inverse\\
Gibbons et al. (2001) & $R$ & 428 & 1.37$\pm 0.04$ & $-0.825\pm 0.01$ & 20\% & inverse \\
Colless et al. (2001) & $R$ & 255 & 1.22$\pm 0.09$ & $-0.84\pm 0.03$ & 20\% &
ML \\ 
Scodeggio (1997) & $I$ & 294 & 1.55$\pm 0.05$ & $-0.80\pm 0.02$ & 22\% & orthogonal \\
Pahre et al. (1998) & $K$ & 251 & 1.53$\pm 0.08$ & $-0.79\pm 0.03$ & 21\% &
orthogonal \\
\tableline
\end{tabular}
\label{fpcoeffslit}
\end{table}

A selection of results from the literature is presented in 
Table~\ref{fpcoeffslit}.  Many of these samples were constructed by 
combining new measurements with previously published photometric and 
velocity dispersion measurements, often made by other authors.  
(Exceptions are J{\o}rgensen et al. 1996, Scodeggio 1997, 
and Colless et al. 2001.)  With respect to previous samples, the SDSS 
sample presented here is both extremely large and homogeneous.  

Notice the relatively large spread in published values of $a$, and the 
fact that $a$ is larger at longer wavelengths.  In contrast, the 
Fundamental Plane we obtain in this paper is remarkably similar in all 
wavebands---although our value of $b$ is consistent with those in the 
literature, the value of $a$ we find in all wavebands is close to the 
largest published values.  In addition, the eigenvectors of our covariance 
matrix satisfy the same relations presented by Saglia et al. (2001).  
Namely, ${\bf\hat v}_1 = {\hat R}_o - a{\hat V} - b{\hat I}_o$, 
${\bf\hat v}_2\approx -{\hat R}_o/b - {\hat V}(1+b^2)/(ab) + {\hat I}_o$ 
and ${\bf\hat v}_3 \approx {\hat R}_o + {\hat I}_o/b$.  
And, when used as a distance indicator, the FP we find is as accurate 
as most of the samples containing more than $\sim 100$ galaxies 
in the literature.  
Unfortunately, at the present time, we have no galaxies in common
with those in any of the samples listed in Table~\ref{fpcoeffslit}, 
so it is difficult to say why our FP coefficients appear to show 
so little dependence on wavelength, or why $a$ is higher than it 
is in the literature.  

The fact that $a\ne 2$ means that the FP is tilted relative to the 
simplest virial theorem prediction $R_o\propto \sigma^2/I_o$.  
One of the assumptions of this simplest prediction is that the 
kinetic energy which enters the virial theorem is proportional 
to the square of the observed central velocity dispersion.  
Busarello et al. (1997) argue that, in fact, the kinetic energy is 
proportional to $\sigma^{1.6}$ rather than to $\sigma^2$.  Since this is 
close to the $\sigma^{1.5}$ scaling we see, it would be interesting to 
see if the kinetic energy scales with $\sigma$ for the galaxies in our 
sample similarly to how it does in Busarello et al.'s sample.  
This requires measurements of the velocity dispersion profiles 
of (a subsample of) the galaxies in our sample, and has yet to 
be done.

Correlations between pairs of observables, such as the 
Faber--Jackson (1976) relation between luminosity and velocity 
dispersion, and the Kormendy (1977) relation between the size and 
the surface brightness, can be thought of as projections of the 
Fundamental Plane.  They are studied in Paper~II.  
The $\kappa$--space projection of Bender, Burstein \& Faber (1992) 
is presented in Section~\ref{kpspace} below.  

\subsection{Residuals and the shape of the FP}\label{fpscat}
Once the FP has been obtained, there are at least two definitions 
of its thickness which are of interest.  
If the FP is to be used as a distance indicator, then the quantity of 
interest is the scatter around the relation in the $R_o$ direction 
only.  On the other hand, if the FP is to be used to constrain models of 
stellar evolution, then one is more interested in the scatter orthogonal 
to the plane.  We discuss both of these below.  

The thickness of the FP is some combination of residuals which are 
intrinsic and those coming from measurement errors.  We would like to 
verify that the thickness is not dominated by measurement errors.  
The residuals from the FP in the different bands are highly correlated; 
a galaxy which scatters above the FP in $g^*$ also scatters above the 
FP in, say, $z^*$.  Although the errors in the photometry in the different 
bands are not completely independent, this suggests that the scatter around 
the FP has a real, intrinsic component.  It is this intrinsic thickness 
which the maximum likelihood analysis is supposed to have estimated.  
The intrinsic scatter may be somewhat smaller than the maximum likelihood 
estimates because there is a contribution to the scatter which comes from 
our assumption that all early-type galaxies are identical when we apply 
the K-correction, for which we have not accounted.  


\begin{figure}
\centering
\epsfxsize=\hsize\epsffile{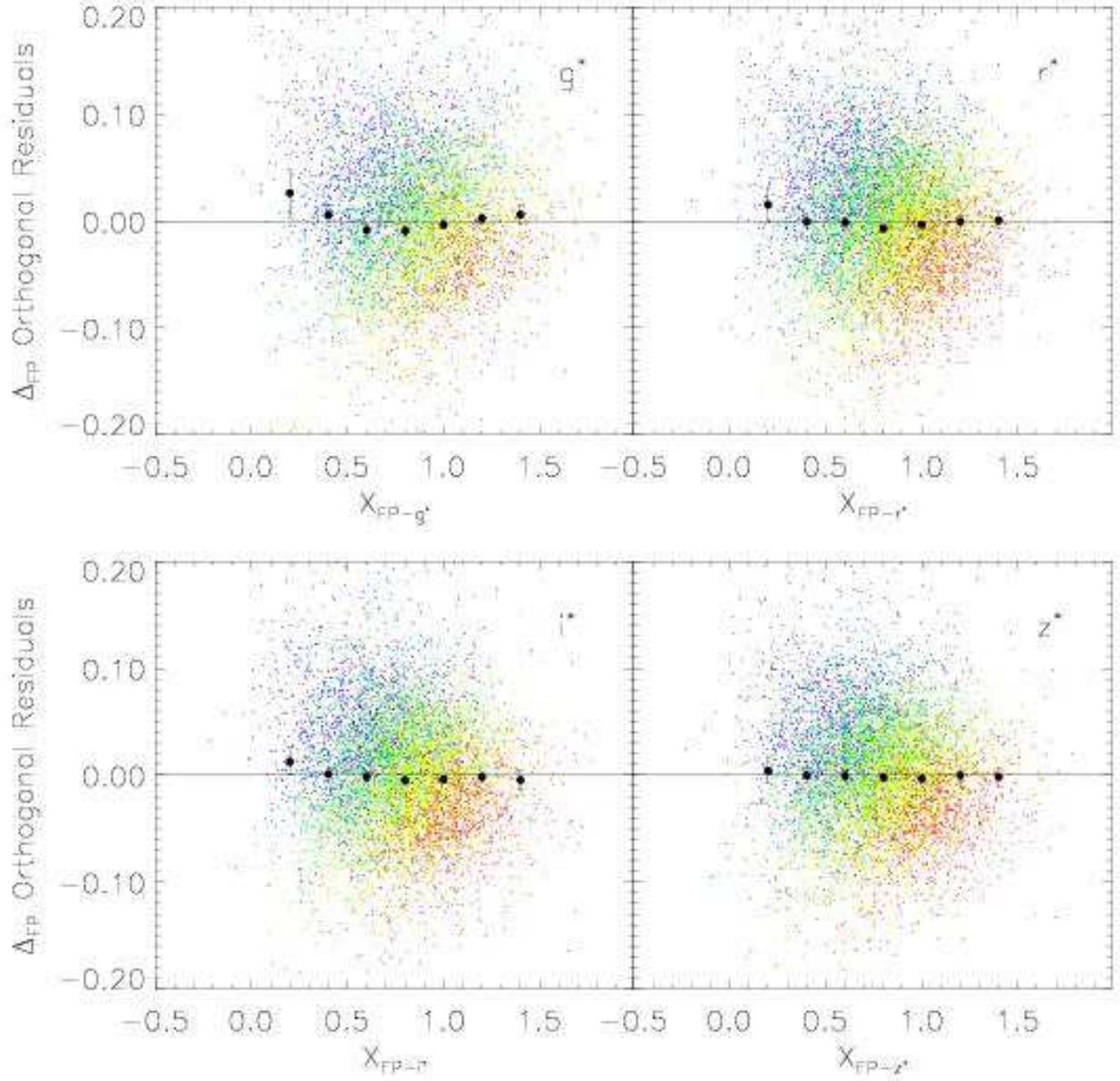}
\caption{Residuals orthogonal to the maximum-likelihood FP fit 
as a function of distance along the fit (the long axis of the plane).  
Error bars show the mean plus and minus three times the error in the 
mean in each bin.  Galaxies with low/high velocity dispersions populate 
the upper-left/lower-right of each panel, but the full sample shows 
little curvature. } 
\label{fig:warped}
\end{figure}

All our estimates of the scatter around the FP show that the FP appears 
to become thicker at shorter wavelengths.  Presumably, this is because 
the light in the redder bands, being less affected by recent 
star-formation and extinction by dust, is a more faithful tracer of 
the dynamical state of the galaxy.  
The orthogonal scatter in our sample, which spans a 
wide range of environments, is comparable to the values given in the 
literature obtained from cluster samples (e.g., Pahre et al. 1998); 
this constrains models of how the stellar populations of early-type 
galaxies depend on environment.  
If the direct fit to the FP is used as a distance indicator, then 
the intrinsic scatter introduces an uncertainty in distance estimates 
of $\sim {\rm ln}(10)\times 0.09\sim$ 20\%.  

Our next step is to check that the FP really is a plane, and not,
for example, a saddle.  To do this, we should show the residuals 
from the orthogonal fit as a function of distance along the 
long axis of the plane.  Specifically, if 
$X\equiv \log_{10}\sigma + (b/a)\log_{10}I_o + (c/a)$, then 
\begin{equation}
X_{FP} \equiv X \sqrt{1+a^2} + (\log_{10}R_o - aX) {a\over\sqrt{1+a^2}}
       =  {X + a\log_{10}R_o \over \sqrt{1+a^2}},
\label{Xfp}
\end{equation}
and we would like to know if the residuals $\Delta_o$ defined earlier  
correlate with $X_{FP}$.  A scatter plot of these residuals versus 
$X_{FP}$ is shown in Figure~\ref{fig:warped} (we have first subtracted 
off the weak evolution in the surface brightnesses).  The symbols 
superimposed on the scatter plot show the mean value of the residuals 
and plus and minus three times the error in the mean, for a few small 
bins along $X_{FP}$.  The figure shows that the FP is reasonably 
flat; it is slightly more warped in the shorter wavelenghts.  

\begin{figure}
\centering
\epsfxsize=\hsize\epsffile{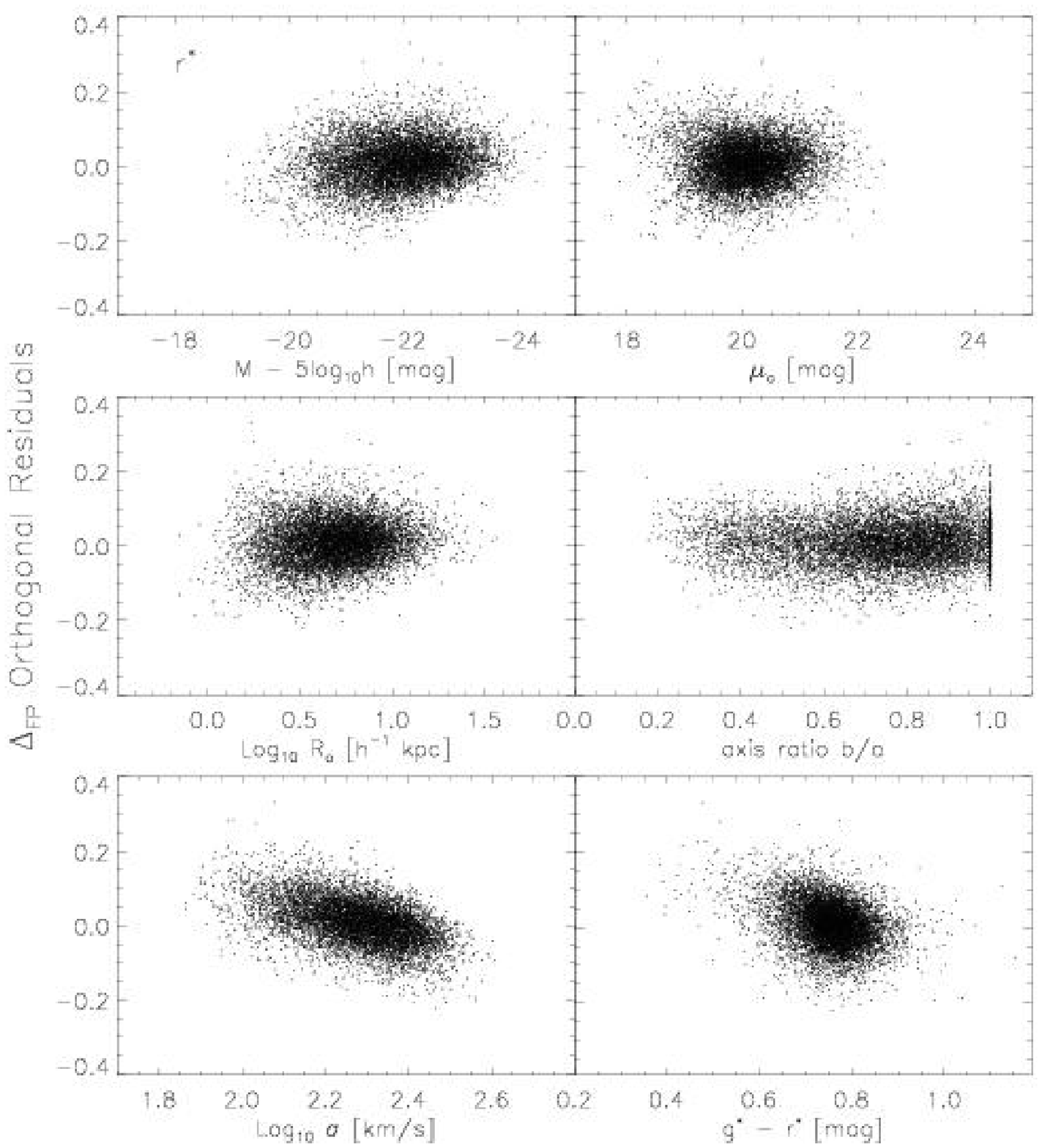}
\caption{Residuals orthogonal to the FP in $r^*$ versus absolute 
magnitude $M$, surface brightness $\mu_o$, 
effective radius $\log_{10} R_o$, axis ratio $b/a$, 
velocity dispersion $\log_{10} \sigma$, and ($g^*-r^*$) color.  
Note the absence of correlation with all parameters other than 
velocity dispersion and color.  }
\label{fig:FPresidR}
\end{figure}

\begin{figure}
\centering
\epsfxsize=\hsize\epsffile{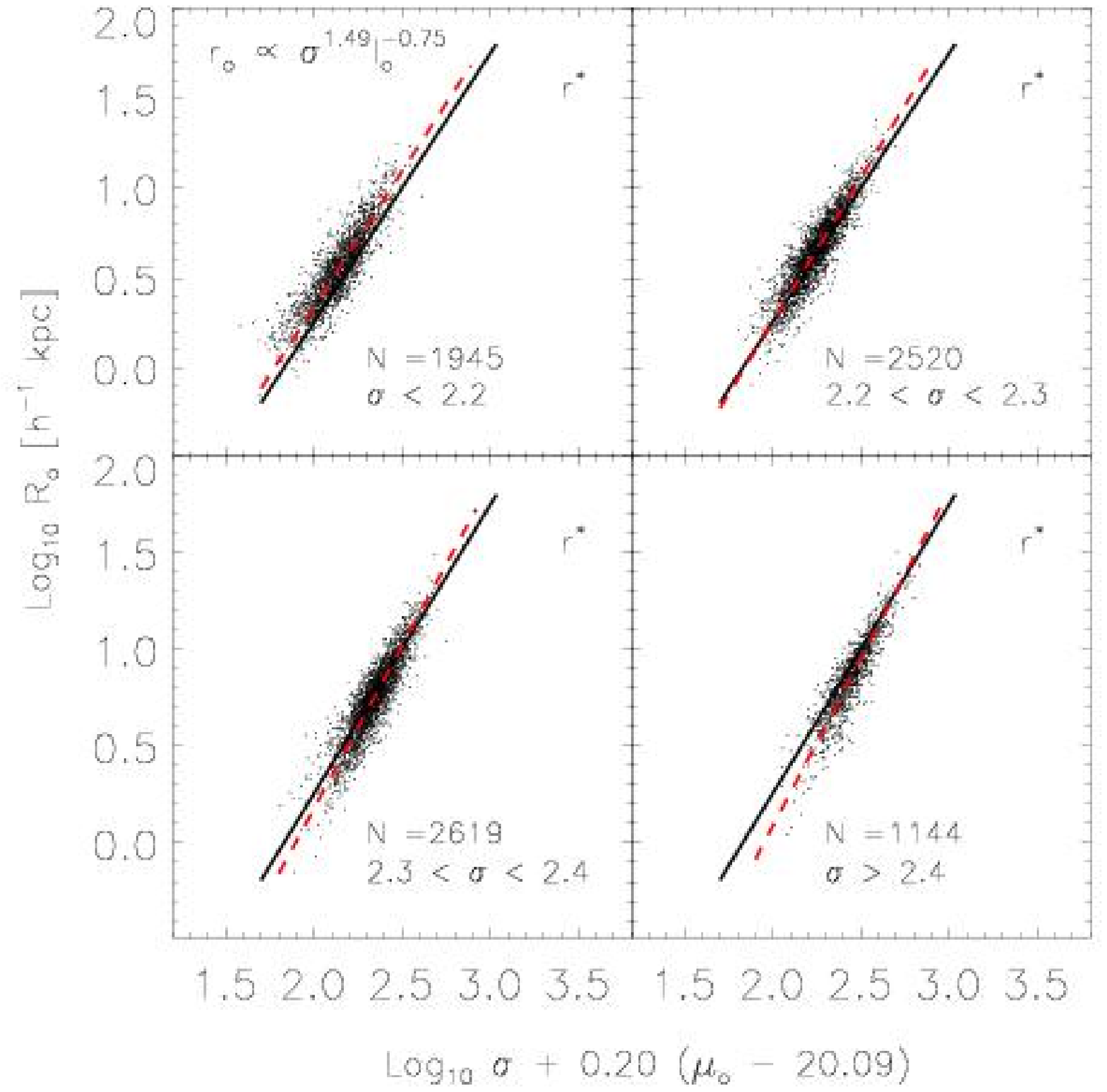}
\caption{The FP in four subsamples defined by velocity dispersion.  
Solid curve (same in all four panels) shows the maximum likelihood 
relation of the parent $r^*$ sample and dashed lines show the 
best-fit, obtained by minimizing the residuals orthogonal to 
the plane, using only the galaxies in each subsample.  The slope of 
the minimization fit increases with increasing velocity dispersion, 
whereas maximum-likelihood fits to the subsamples, which account for 
the cut on $\sigma$, give the same slope as for the full sample.  } 
\label{fig:FPsigR}
\end{figure}

Given that the FP is not significantly warped, we would like to know 
if deviations from the Plane correlate with any of the three physical 
parameters used to define it.  When the plane is defined by minimizing 
with respect to $\log_{10} R_o$, there is little if any correlation of 
the residuals with absolute magnitude, surface brightness, 
effective radius, axis-ratio, velocity dispersion, or color so we 
have chosen to not present them here.  
Instead, Figure~\ref{fig:FPresidR} shows the result of plotting the 
residuals orthogonal to the plane when the plane is defined by the 
orthogonal fit.  The residuals show no correlation with 
$M$, $\mu_o$, $\log_{10} R_o$, or axis ratio (we have subtracted the 
weak evolution in $M$ and $\mu_o$ when making the scatter plots).  
The residuals are anti-correlated with $\log_{10}\sigma$ and 
slightly less anti-correlated with $(g^*-r^*)$ color.  The correlation 
with color is due to the fact that velocity dispersion and color are 
tightly correlated (this correlation is studied in more detail in Paper~IV).  
The correlation with velocity dispersion is not a selection effect, 
nor is it associated with evolution; we see a similar trend with 
velocity dispersion in both the complete and the magnitude-limited 
simulated catalogs.  

Figure~\ref{fig:FPsigR} shows why this happens.  The four panels show 
the FP in four subsamples of the full $r^*$ sample, divided according 
to velocity dispersion.  
Notice how the different scatter plots in Figure~\ref{fig:FPsigR} 
show sharp cut-offs approximately perpendicular to the x-axis:  
lines of constant $\sigma$ are approximately perpendicular to the 
x-axis.  Whereas the direct fit is not affected by a cut-off which 
is perpendicular to the x-axis, the orthogonal fit is.  Hence, the 
residuals with respect to the orthogonal fit show a correlation with 
velocity dispersion, whereas those from the direct fit do not.  
(Indeed, by using the coefficients provided in Tables~\ref{MLcov} 
and~\ref{fpcoeffs}, and the definition of the residuals $\Delta_1$ 
and $\Delta_o$, one can show that 
$\langle \Delta_1|\log_{10}\sigma \rangle$ is proportional to 
$\log_{10}\sigma$, with a constant of proportionality which is 
close to zero when the parameters for the direct fit are inserted.  
However, when the parameters for the orthogonal 
fit are used, then the slope of the 
$\langle\Delta_o|\log_{10}\sigma\rangle$ versus $\log_{10}\sigma$ 
relation is significantly different from zero.)  

To illustrate, the solid curves in Figure~\ref{fig:FPsigR} (the same in 
each panel) show the maximum likelihood FP for the full sample.  The 
dashed curves show the FP, determined by using the $\chi^2-$method to 
minimize the residuals orthogonal to the plane, in various subsamples 
defined by velocity dispersion.  
The panels for larger velocity dispersions show steeper relations.  
Evidence for a steepening of the relation with increasing velocity 
dispersion was seen by J{\o}rgensen et al. (1996).  Their sample was 
considerably smaller than ours, and so they ruled the trend they saw 
as only marginal.  Our much larger sample shows this trend clearly.  
We have already argued that this steepening is an artifact of the 
fact that lines of constant velocity dispersion are perpendicular to the 
$x$-axis.  The maximum-likelihood fit to the subsamples is virtually the 
same as that for the full sample, provided we include the correct velocity 
dispersion cuts in the normalization of the likelihood.  
In other words, the maximum-likelihood fit is able to account for the 
bias introduced by making a cut in velocity dispersion as well as 
apparent magnitude.  

\subsection{The mass-to-light ratio}\label{m2l}
The Fundamental Plane is sometimes used to infer how the mass-to-light 
ratio depends on different observed or physical parameters.  
For example, the scaling required by the virial theorem, 
$M_o\propto R_o\sigma^2$, 
combined with the assumption that the mass-to-light ratio scales as 
$M_o/L\propto M_o^\gamma$ yields a Fundamental Plane like relation 
of the form:
\begin{equation}
R_o\propto \sigma^{2(1-\gamma)/(1+\gamma)}I_o^{-1/(1+\gamma)}.
\end{equation}
The observed Fundamental Plane is $R_o\propto \sigma^a\,I_o^b$.  
If the relation above is to describe the observations, then 
$\gamma$ must simultaneously satisfy two relations:  
$\gamma = (2-a)/(2+a)$, and $\gamma=-(1+b)/b$.  
The values of $b$ in the literature are all about $-0.8$; setting 
$\gamma$ equal to the value required by $b$ and then writing $a$ in 
terms of $b$ gives $a = -2(1+2b)$.  Most of the values of $a$ and $b$ 
in the shorter wavebands reported in the literature (see, e.g., 
Table~\ref{fpcoeffslit}) are consistent with this scaling, whereas the 
higher values of $a$ found at longer wavelengths are not.  Although the 
direct fits to our sample have small values of $a$, the orthogonal fits 
give high values in all four bands.  These fits do not support the 
assumption that $M_o/L$ can be parametrized as a function of $M_o$ alone.  

\begin{figure}
\centering
\epsfxsize=\hsize\epsffile{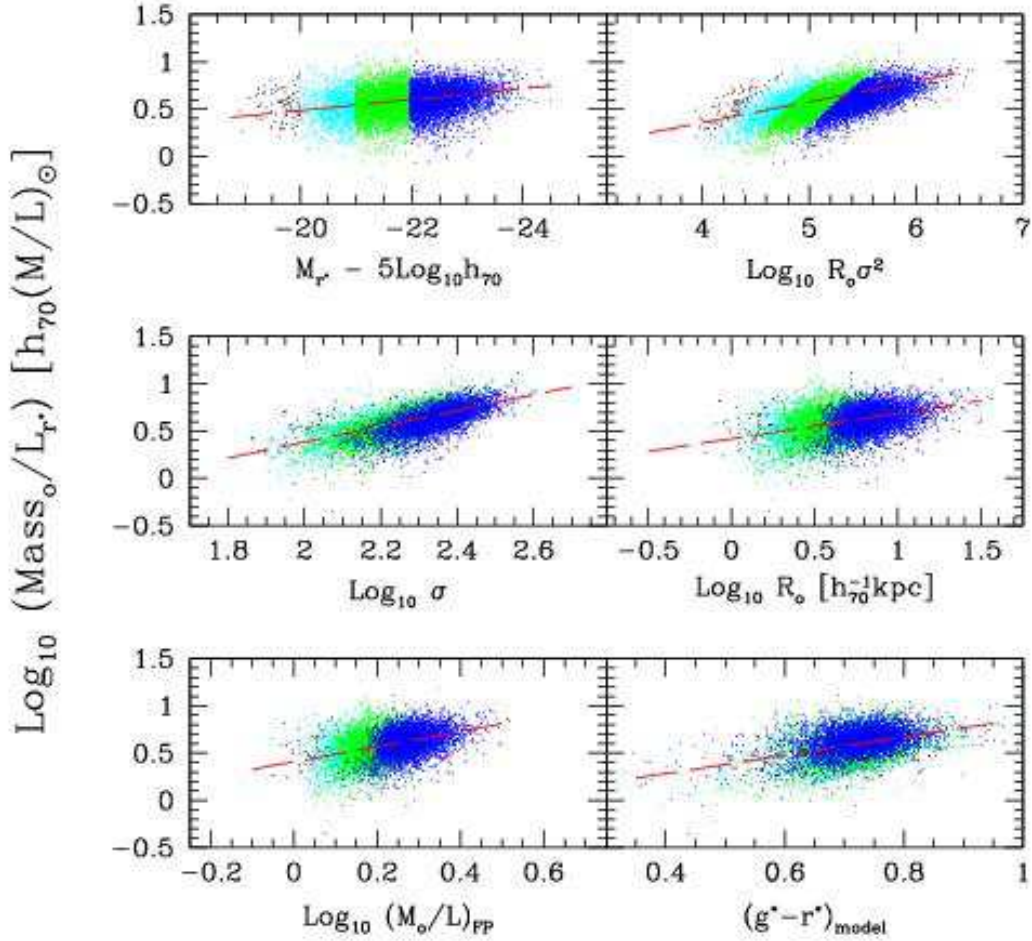}
\vspace{1cm}
\caption{Ratio of effective mass $R_o\sigma^2$ to effective luminosity 
$(L/2)$ as a function of luminosity (top left), mass (top right), 
velocity dispersion (middle left), surface brightness (middle right), 
the combination of velocity dispersion and size suggested by the 
Fundamental Plane (bottom left), and color (bottom right).  
Notice the substantial scatter around the best fit linear relation 
in the bottom left panel, the slope of which is shallower than unity.}
\label{fig:m2l}
\end{figure}

Another way to phrase this is to note that, when combined with the 
virial theorem requirement that $(M_o/L)\propto \sigma^2/(R_oI_o)$, 
the Fundamental Plane relation $R_o\propto \sigma^aI_o^b$ yields 
\begin{equation}
\left({M_o\over L}\right)_{\rm FP} \propto\sigma^{2 + a/b}\,R_o^{-(1+b)/b} 
\label{m2lfp}
\end{equation}
(e.g. J{\o}rgensen et al. 1996; Kelson et al. 2000).
The quantity on the right hand side is the mass-to-light ratio 
`predicted' by the Fundamental Plane, if $\sigma$ and $R_o$ 
are given, and the scatter in the Fundamental Plane is ignored.  
This is a function of $M_o$ alone only if $a = -2(1+2b)$.  
Our orthogonal fit coefficients $a$ and $b$ are not related in this 
way.  Rather, for our Fundamental Plane, the dependence on 
$\sigma$ in equation~(\ref{m2lfp}) cancels out almost exactly:  
to a very good approximation, we find 
 $(M_o/L)_{\rm FP}\propto R_o^{-(1+b)/b}\propto R_o^{0.33}$.  
Alternatively, a little algebra shows that mass to light ratio is 
determined by the combination $(\sigma^2/I_o)^{0.25}$.  
Whether there is a simple physical reason for this is an open question. 

In contrast to the predicted ratio, $(M_o/L)_{\rm FP}$, the combination 
$R_o\sigma^2/L$ is the `observed' mass-to-light ratio.  The ratio of the 
observed value to the FP prediction of equation~(\ref{m2lfp}) is 
$(R_o/I_o^b\sigma^a)^{1/b}$.  The scatter in the logarithm of this ratio 
is $1/b$ times the scatter in Fundamental Plane in the direction of $R_o$ 
(i.e., it is the scatter in the quantity we called $\Delta_1$ in the 
previous subsections, divided by $b$).  
Inserting the values from Table~\ref{fpcoeffs} shows that if the values of 
$\sigma$ and the effective radius in $r^*$ are used to predict the values 
of the mass-to-light ratio in $r^*$, then the uncertainty in the predicted 
ratio is 26\%.  This is larger than the values quoted in the literature 
for early-type galaxies in clusters (e.g., J{\o}rgensen et al. 1996; 
Kelson et al. 2000).  

Unfortunately, this is somewhat confusing terminology, because 
the two mass-to-light ratios are not proportional to each other.  
This can be seen by using the maximum-likelihood results of 
Table~\ref{MLcov} to compute the mean of the observed mass to light 
ratio $R_o\sigma^2/L$ at fixed predicted $(M/L)_{\rm FP}$, 
or simply by plotting the two quantities against one another.  
Figure~\ref{fig:m2l} shows how $R_o\sigma^2/L$ correlates with 
luminosity, mass $R_o\sigma^2$, velocity dispersion, surface brightness, 
the ratio predicted by the Fundamental Plane, and color.  The different 
panels show obvious correlations; the maximum likelihood predictions 
for these correlations can be derived from the coefficients in 
Table~\ref{MLcov}:  
$(R_o\sigma^2/L)\propto L^{0.14\pm 0.02}$, 
$(R_o\sigma^2/L)\propto (R_o\sigma^2)^{0.22\pm 0.05}$, 
$(R_o\sigma^2/L)\propto \sigma^{0.84\pm 0.1}$, and 
$(R_o\sigma^2/L)\propto R_o^{0.27\pm 0.06}$.  
These are shown as dashed lines in the top four panels.  
A linear fit to the scatter plot in the bottom left panel gives 
$(R_o\sigma^2/L)\propto (M/L)_{\rm FP}^{0.80\pm 0.05}$, with an 
rms scatter around the fit of 0.14:  the ratio predicted by the 
Fundamental Plane is not proportional to the observed ratio.  
A scatter plot of $(M/L)_{\rm FP}$ against all these quantities is 
tighter, of course (recall the scatter around the FP has been removed), 
although some of the slopes are significantly different.  
For example, $(M/L)_{\rm FP}\propto L^{0.16\pm 0.04}$, 
$(M/L)_{\rm FP}\propto (R_o\sigma^2)^{0.13\pm 0.03}$, and
$(M/L)_{\rm FP}\propto \sigma^{0.21\pm 0.03}$:  
the `observed' and `predicted' slopes of the mass-to-light ratio 
versus $\sigma$ relations are very different.   
For this reason, one should be careful in interpretting what 
is meant by the `predicted' mass-to-light ratio.  Our own view 
is that the observed ratio, $R_o\sigma^2/L$ is to be preferred, 
as it is directly related to observables, and is independent of the 
fitting procedure used to fit the Fundamental Plane.

\subsection{The Fundamental Plane:  Evidence for evolution?}\label{evolve}
\begin{figure}
\centering
\epsfxsize=\hsize\epsffile{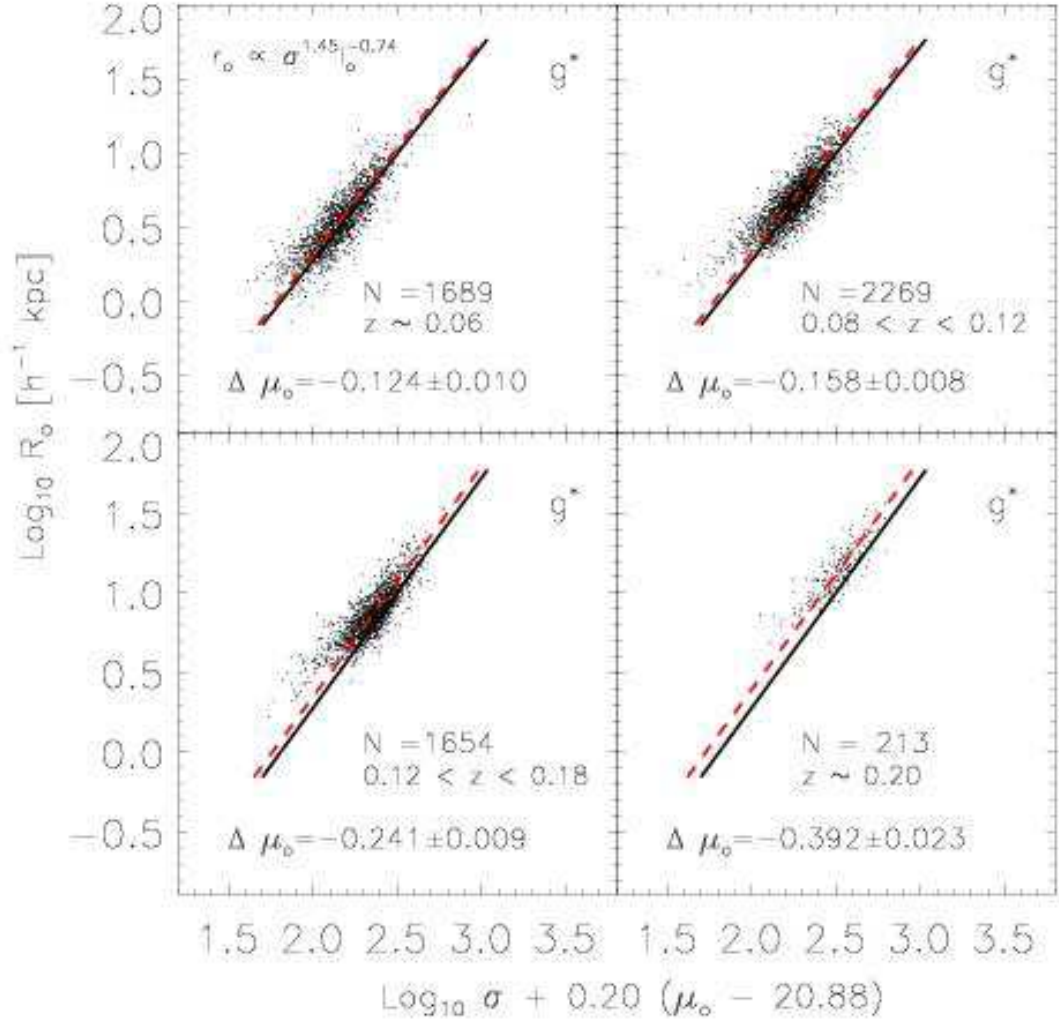}
\vspace{1cm}
\caption{The $g^*$ FP in four redshift bins. The slope of the FP is 
fixed to that at zero redshift; only the zero-point is allowed 
to vary.  The zero-point shifts systematically with redshift.  
The same plot for $r^*$ shows similar but smaller shifts.  }
\label{fig:FPzG}
\end{figure}

The Fundamental Plane is sometimes used to test for evolution.  
This is done by plotting $R_o$ versus the combination of $\mu_o$ and 
$\sigma$ which defines the Fundamental Plane at low redshift, and then 
seeing if the high-redshift population traces the same locus as the low 
redshift population.  Figure~\ref{fig:FPzG} shows this test for our 
$g^*$ band sample:  solid lines (same in each panel) show the relation 
which fits the zero-redshift sample; dashed lines show a line with the 
same slope which best-fits the higher redshift sample.  The population at 
higher redshift is displaced slightly to the left of the low redshift 
population; the text in the bottom of each panel shows this shift, 
expressed as a change in the surface brightness $\mu_o$.  
The plot appears to show that, on average, the higher redshift galaxies 
are brighter, with the brightening scaling approximately as 
$\Delta\mu_o \approx -2z$.  

\begin{figure}
\centering
\epsfxsize=\hsize\epsffile{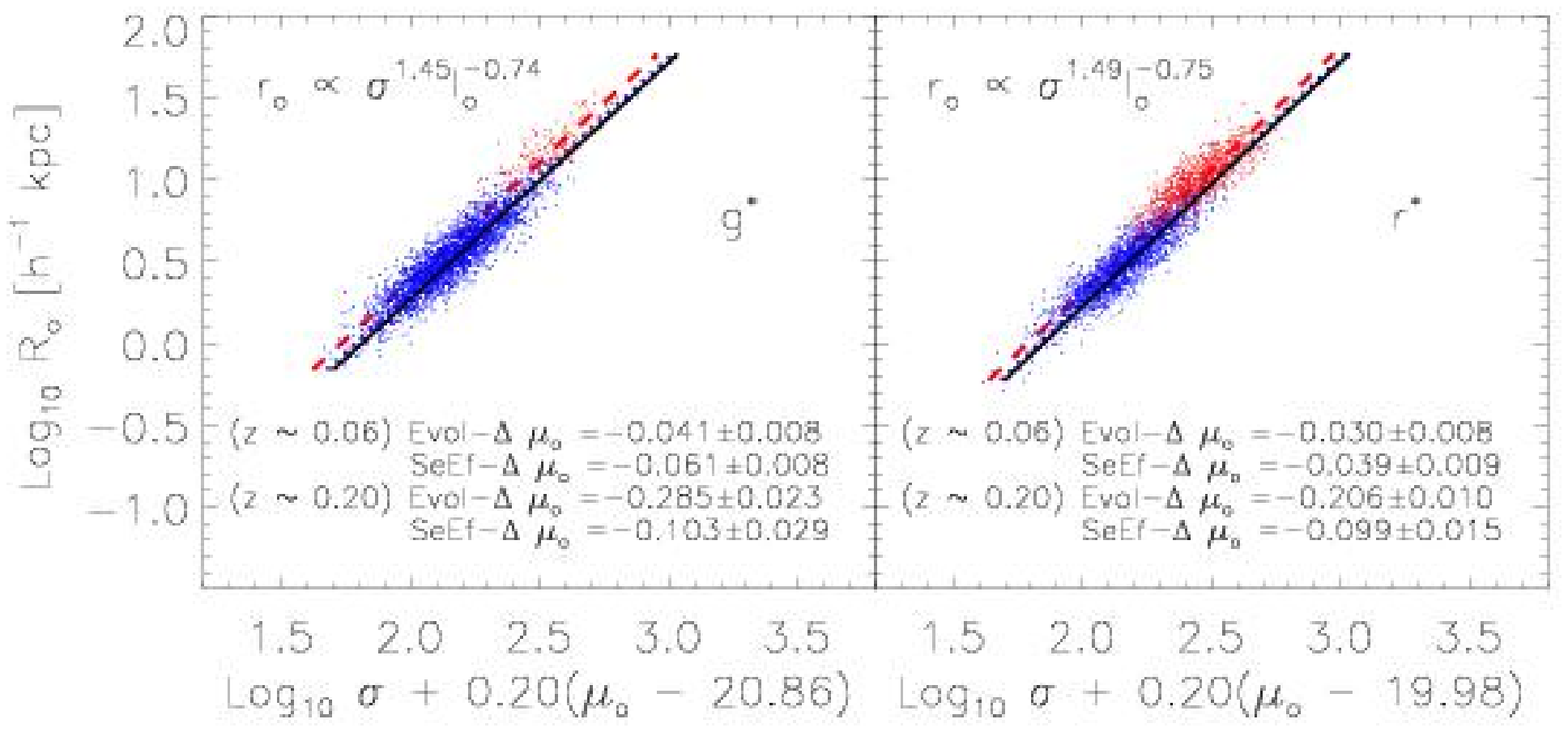}
\vspace{0cm}
\caption{The FP in the $g^*$ (left panel) and $r^*$ (right panel)
magnitude-limited mock catalogs.  Solid line shows the FP at $z=0$.  
Dotted and dashed lines show fits using a low and high redshift 
subsample only.  For these fits, the slope of the FP is required to be 
the same as the solid line; only the zero-point is allowed to vary.  
The shift seen in the complete simulations is labeled 
`Evol$-\Delta\mu_o$', whereas the shift seen in the magnitude limited 
simulations is the sum of this and the quantity labeled 
`SeEf$-\Delta\mu_o$'.  This sum is similar to the shift seen in 
the SDSS data, suggesting that selection effects are not negligble.}
\label{fig:FPsimz}
\end{figure}

How much of this apparent brightening is really due to evolution, and 
how much is an artifact of the fact that our sample is magnitude limited?  
To address this, we generated complete and magnitude limited mock galaxy 
catalogs as described in Appendix~A of Paper~II, and then performed the same 
test for evolution.  Comparing the shifts in the two simulations allows 
us to estimate how much of the shift is due to the selection effect.  
Figure~\ref{fig:FPsimz} shows the results in our simulated $g^*$ (left) 
and $r^*$ (right) catalogs.  The solid lines in each panel show the 
zero-redshift relation, and the dotted and dashed lines show lines of the 
same slope which best-fit the points at low and high redshift, respectively.  
The text in the bottom shows how much of the shift in $\mu_o$ is due 
to the magnitude limit, and how much to evolution.  The sum of the two 
contributions is the total shift seen in the magnitude limited simulations.
Notice that this sum is similar to that seen in the data 
(Figure~\ref{fig:FPzG}), both at low and high redshifts, suggesting 
that our simulations describe the varying roles played by evolution and 
selection effects accurately.  Since the parameters of the simulations 
were set by the maximum likelihood analysis, we conclude that the 
likelihood analysis of the evolution in luminosities is reasonably 
accurate ($\Delta\mu_o\approx -1.15z$) in $g^*$, but we note that this 
evolution is less than one would have infered if selection effects 
were ignored ($\Delta\mu_o\approx -2z$).  

The importance of selection effects in our sample has implications 
for another way in which studies of evolution are presented.  
If galaxies do not evolve, then the FP can be used to define a standard 
candle, so the test checks if residuals from the FP in the direction of 
the surface-brightness variable, when plotted versus redshift, follow 
Tolman's $(1+z)^4$ cosmological dimming law.  
If Friedmann-Robertson-Walker models are correct, then departures from 
this $(1+z)^4$ dimming trend can be used to test for evolution.  
This can be done if one assumes that the main effect of evolution is 
to change the luminosities of galaxies.  If so, then evolution will 
show up as a tendency for the residuals from the FP, in the $\mu$ 
direction, to drift away from the $(1+z)^4$ dimming 
(e.g., Sandage \& Perelmuter 1990; Pahre, Djorgovski \& de Carvalho 1996).  

\begin{figure}
\centering
\epsfxsize=\hsize\epsffile{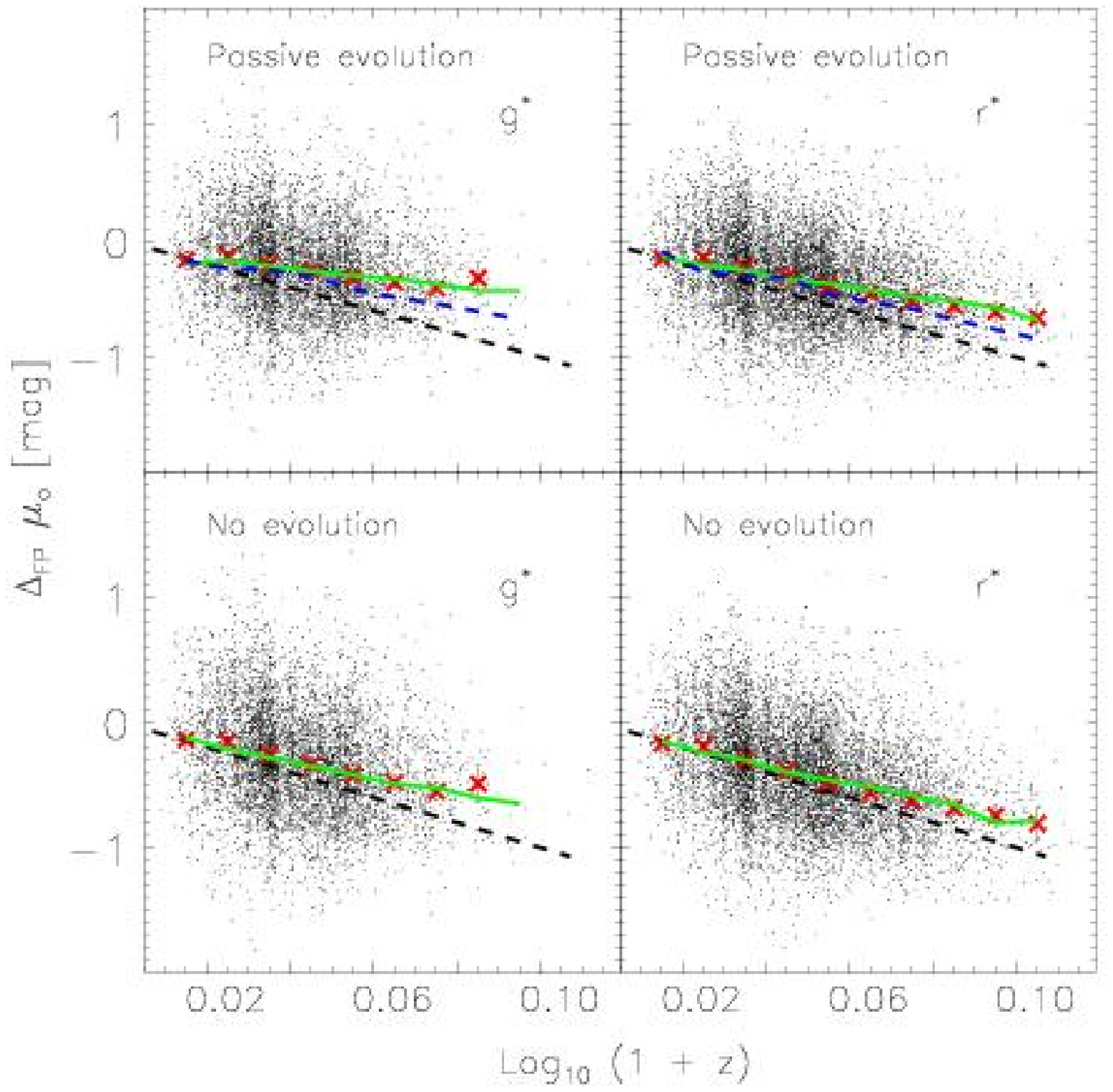}
\caption{Residuals of the zero-redshift FP with respect to the 
surface brightness, before correcting for cosmological dimming, 
versus redshift in the four bands.  Lowest dashed line in all panels 
shows the $(1+z)^4$ dimming expected if there is no evolution.  Solid 
curves in top panels show the same measurement in mock simulations of 
a magnitude limited sample of a passively evolving population.  Dashed 
lines in between show the actual evolution in surface brightness---the 
difference between these and the solid curves is an artifact of the 
magnitude limit.  Bottom panel shows the same test applied using the 
parameters of the Fundamental Plane which best describes the data if 
there is required to be no evolution whatsoever.  Solid lines show what 
one would observe in a magnitude limited sample of such a population.  
In this case, the entire trend away from the $(1+z)^4$ dimming is a 
selection effect.  Note how, once the magnitude limit has been applied, 
both the evolving (top) and non-evolving populations (bottom) appear 
very similar to our observed sample.  }
\label{fig:FPresidmuz}
\end{figure}

Figure~\ref{fig:FPresidmuz} shows this trend in our dataset.  
The lowest dashed lines in all panels show the expected $(1+z)^4$ 
dimming; panels on the left/right show results in $g^*$/$r^*$.  
Consider the top two panels first.  The points show residuals with respect 
to the zero-redshift Fundamental Plane in our sample.  The crosses show the 
median residual in a small redshift bin.  The galaxies do not quite follow 
the expected $(1+z)^4$ dimming.  The similarity to the $(1+z)^4$ dimming 
argues in favour of standard cosmological models, whereas the small 
difference from the expected trend is sometimes interpretted as evidence 
for evolution (e.g., J{\o}rgensen et al. 1999; van Dokkum et al. 1998, 2001; 
Treu et al. 1999, 2001a,b).  

Of course, to correctly quantify this evolution, we must account for 
selection effects.  The dashed lines which lie between the $(1+z)^4$ 
scaling and the data (i.e., the crosses) show how the surface brightness 
should scale if there were passive evolution of the form suggested 
by the maximum likelihood analysis, but there were no magnitude 
limit.  That is, if $M_*(z)=M_*(0)-Qz$, then the surface brightnesses 
should scale as $(1+z)^{4-0.92\,Q}$.
The solid curves show the result of making the measurement in simulated 
magnitude limited catalogs which include this passive evolution.  
Notice how different these solid curves are from the dashed curves (they 
imply $Q$ about twice the correct value), but note how similar they are 
to the data.  This shows that about half of the evolution one would 
naively have infered from such a plot is a consequence of the magnitude 
limit.  

To further emphasize the strength of this effect, we constructed simulations 
in which there was no evolution whatsoever.  We did this by first making 
maximum likelihood estimates of the joint luminosity, size and velocity 
dispersion distribution in which no evolution was allowed.  (For the 
reasons discussed earlier, the associated no-evolution Fundamental Plane 
coefficient $a$ is steeper by about 10\%.)  This was then used to generate 
mock catalogs in which there is no evolution.  The crosses in the bottom 
panels show the result of repeating the same procedure as in the top panels, 
but now using the parameters of the no-evolution Fundamental Plane, and the 
solid line shows the measurement in the no-evolution simulations in which, 
by construction, the population of galaxies at all redshifts is identical.  
Therefore, the shifts from the $(1+z)^4$ dimming we see in the magnitude 
limited no-evolution catalogs (solid curves in bottom panels) are entirely 
due to the magnitude limit.  Notice how similar the solid lines from our 
no-evolution simulations are to the actual data.  If we believed there 
really were no evolution, then the results shown in the bottom panel would 
lead us to conclude that much of the trend away from the $(1+z)^4$ dimming 
is a selection effect---it is not evidence for evolution.  

(The fact that we were able to find a non-evolving population which mimics 
the observations so well suggests that the population of early-type 
galaxies at the median redshift of our sample must be rather similar 
to the population at lower and at higher redshifts.  This, in turn, can 
constrain models of when the stars in these galaxies must have formed.)

We view our no-evolution simulations as a warning about the accuracy of 
this particular test of evolution.  If the evolution is weak, then it 
appears that the results of this test depend critically on how the catalog 
was selected, and on what one uses as the fiducial Fundamental Plane.  
To make this second point, we followed the procedure adopted by many other 
recent publications.  Namely, we assumed that the zero-redshift Fundamental 
Plane has the shape reported by J{\o}rgensen et al. (1996) for Coma, for 
which $a$ is about 15\% smaller than what we find in $g^*$.  If no account 
is taken of selection effects, then the inferred evolution in $\mu_o$ 
results in a value of $Q$ which is about a factor of four times larger 
than the one we report in Table~\ref{MLcov}!  

Our results indicate that inferences about evolution which are based on 
this test depend uncomfortably strongly on the strength of selection 
effects, and on what one assumes for the fiducial shape of the Fundamental 
Plane.  In this respect, our findings about the role of, and the need to 
account for selection effects are consistent with those reported by 
Simard et al. (1999).  While we believe we have strong evidence that 
the early-type population is evolving, we do not believe that the 
strongest evidence of this evolution comes from either of the tests 
presented in this subsection.  Nevertheless,
it is reassuring that the evolution we see from these Fundamental Plane 
tests is consistent with that which we estimated using the likelihood 
analysis in Paper~II, and is also consistent with what we use to make our 
K-corrections.  Namely, a passively evolving population which formed 
the bulk of its stars about 9~Gyrs ago appears to provide a reasonable 
description of the evolution of the surface brightnesses in our 
sample.  

\subsection{The Fundamental Plane:  Dependence on environment}\label{environ}

This section is devoted to a study of if and how the properties of 
early-type galaxies depend on environment.  Paper~I describes 
our working definition of environment---essentially, we use the 
number of galaxies which are nearby in color-, angular- and 
redshift-space as an indication of the local density.  Our procedure 
for assigning neighbours is least secure in the lowest redshift bin 
(typically $z\le 0.08$).  


\begin{figure}
\centering
\epsfxsize=\hsize\epsffile{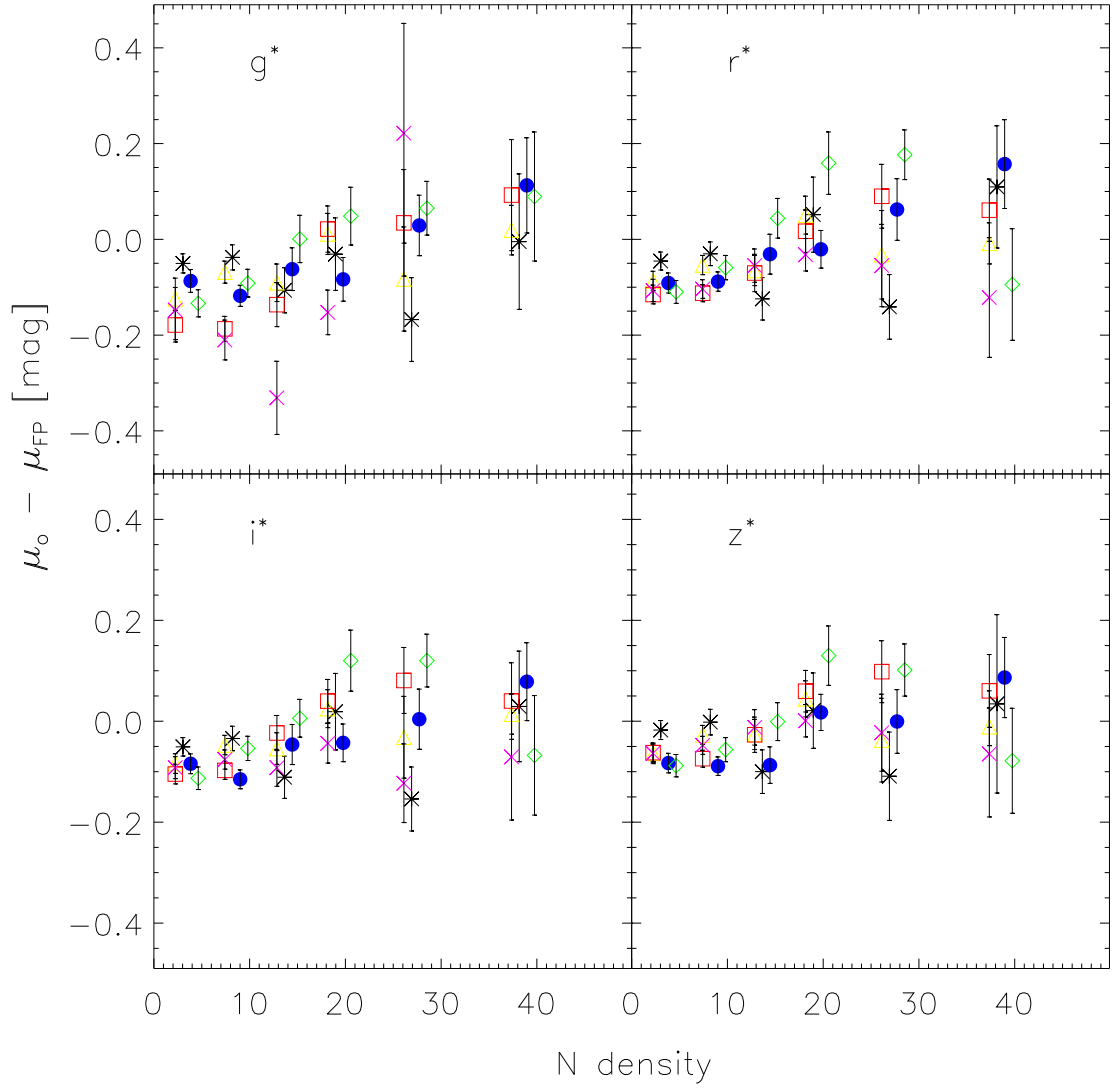}
\vspace{-1.cm}
\caption{Residuals from the FP as a function of number of nearby 
neighbours.  Stars, circles, diamonds, triangles, squares and crosses show 
averages over galaxies in the redshift ranges $z\le 0.075$, $0.075< z\le 0.1$, 
$0.1< z\le 0.12$, $0.12< z\le 0.14$, $0.14< z\le 0.18$ and $z>0.18$.}  
\label{FPenviron}
\end{figure}

Paper~I shows that when the number of near neighbours is small, 
the luminosities, sizes and velocity dispersions all increase slightly 
as the local density increases, whereas the surface brightnesses 
decrease slightly, although all these trends are very weak.  
A more efficient way of seeing if the properties of galaxies depend 
on environment is to show the residuals from the Fundamental Plane.  
As we argue below, this efficiency comes at a cost:  if the residuals 
correlate with environment, it is 
difficult to decide if the correlation is due to changes in luminosity, 
size or velocity dispersion.  

Figure~\ref{FPenviron} shows the differences between galaxy surface 
brightnesses and those predicted by the zero-redshift maximum likelihood 
FP given their sizes and velocity dispersions, as a function of local 
density.  Stars, circles, diamonds, triangles, squares and crosses show 
averages over galaxies in the redshift ranges $z\le 0.075$, $0.075< z\le 0.1$, 
$0.1< z\le 0.12$, $0.12< z\le 0.14$, $0.14< z\le 0.18$ and $z>0.18$.  
Error bars show the error in determining the mean.  (For clarity, the 
symbols have been offset slightly from each other.)  
The plot shows that the residuals depend on redshift---we have already 
argued that this is a combination of evolution and selection effects.  
Notice, that in all redshift bins, the residuals tend to 
increase as local density increases.  This suggests that the mean  
residual from the Fundamental Plane depends on environment.  
If the offset in surface brightness is interpretted as evidence that 
galaxies in denser regions are slightly less luminous than their 
counterparts in less dense regions, then this might be evidence 
that they formed at higher redshift.  While this is a reasonable 
conclusion, we should be cautious:  because 
$\mu_o - \mu_{\rm FP}(R_o,\sigma) = -\Delta_1/b$, what we 
have really found is that the residuals in the direction of $R_o$ 
correlate with environment.  
Because $\sigma - \sigma_{\rm FP}(R_o,\mu_o) = -\Delta_1/a$, 
we might also have concluded that the velocity dispersions of galaxies 
in dense regions are systematically different from those of galaxies 
which have the same sizes and luminosities but are in the field.  
For similar reasons, a plot of the mean residual orthogonal to the 
plane shows a dependence on environment. (However, the rms scatter 
in the orthogonal direction of the residuals around the mean 
residual in each density bin, when plotted as a function of density, 
shows no trend.)  
Thus, while the Fundamental Plane suggests that the properties of 
galaxies depend on environment, it does not say how.

\subsection{The $\kappa$-space projection}\label{kpspace}
Bender et al. (1992) suggested three simple combinations of the three 
observables:
\begin{eqnarray}
\kappa_1 &=& {\log_{10} (R_o\,\sigma^2)\over\sqrt{2}} ,\nonumber\\
\kappa_2 &=& {\log_{10} (I_o^2\sigma^2/R_o)\over\sqrt{6}},
\qquad {\rm and}\nonumber\\
\kappa_3 &=& {\log_{10} (I_o^{-1}\sigma^2/R_o)\over\sqrt{3}},
\end{eqnarray}
which, they argued, correspond approximately to the FP viewed 
face-on ($\kappa_2$--$\kappa_1$), and the two edge-on projections 
($\kappa_3$--$\kappa_1$ and $\kappa_3$--$\kappa_2$).  
They also argued that their parametrization was simply related to the 
underlying physical variables.  For example, $\kappa_1\propto$ mass 
and $\kappa_3\propto$ the mass-to-light ratio.  
The $\kappa_1$--$\kappa_2$ projection would view the FP face on 
if $R_o\propto \sigma^2/I_o$; recall that we found 
$R_o\propto (\sigma^2/I_o)^{0.75}$.  

\begin{figure}
\centering
\epsfxsize=\hsize\epsffile{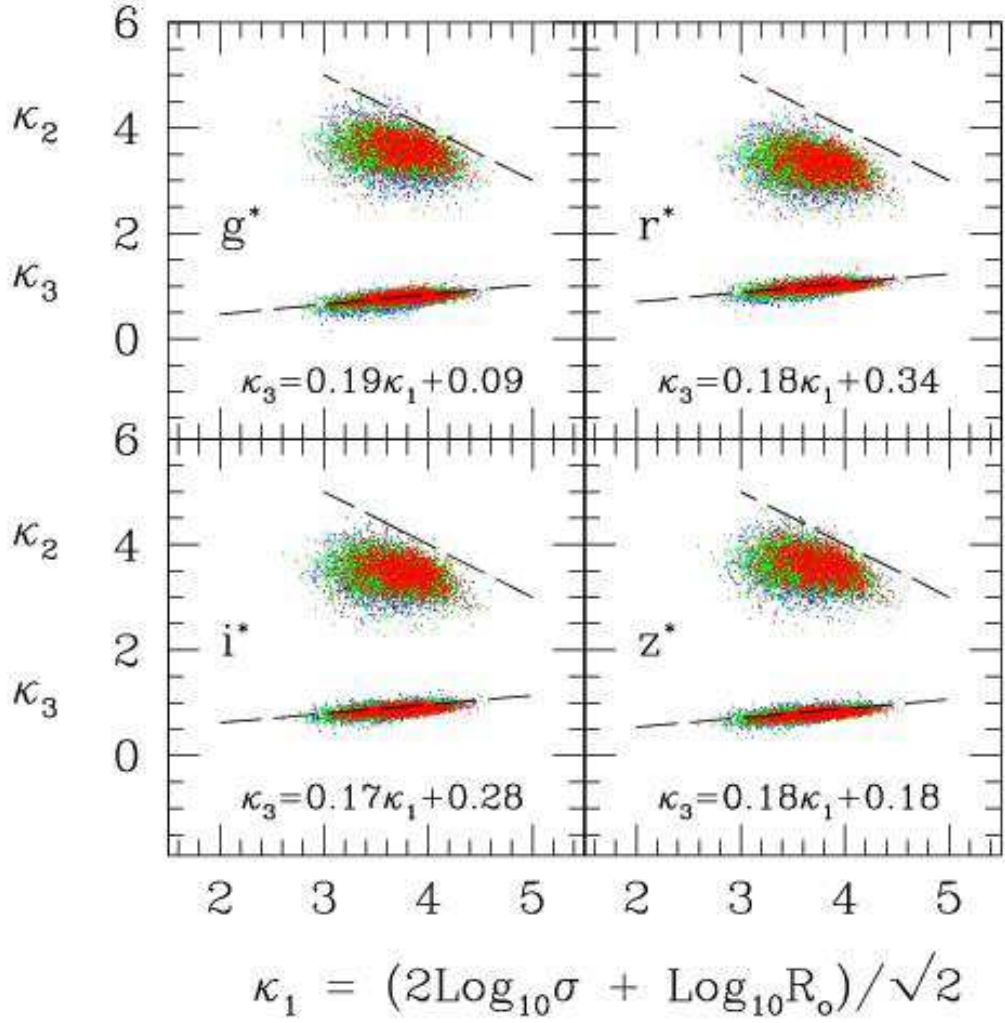}
\vspace{0.cm}
\caption{The early-type sample in the four SDSS bands viewed 
in the $\kappa$-space projection of Bender et al. (1992).  
The dashed line in the upper right corner of each panel shows 
$\kappa_1+\kappa_2=8$, what Burstein et al. (1997) termed the 
`zone of avoidance'.  }
\label{fpk4}
\end{figure}

Bender et al.'s choice of parameters was criticized by Pahre et al. (1998) 
on the grounds that if the effective radius $R_o$ is a function 
of wavelength, then the `mass' becomes a function of wavelength, which 
is unphysical.  On average, the effective radii of the galaxies in our 
sample do increase with decreasing wavelength (Paper~I), so one 
might conclude that Pahre et al.'s objections are valid.  
However, recall that we do not use the measured velocity dispersion 
directly; rather, $\sigma$ represents the value the dispersion would 
have had at some fixed fraction of $R_o$.  If $R_o$ depends on 
wavelength, and we wish to measure the velocity dispersion at a 
fixed fraction of $R_o$, then one might argue that we should also 
correct the measured velocity dispersion differently in the different 
bands.  The velocity dispersion decreases with increasing radius.  
So, if $R_o$ is larger in the blue band than the red, then the 
associated velocity dispersion we should use in the blue band should 
be smaller than in the red.  One might imagine that the combination 
$R_o\sigma^2$ remains approximately constant after all.  
Whether or not it does, we have chosen to present the SDSS early-type 
sample in the $\kappa$-space projection introduced by Bender et al.  

Figure~\ref{fpk4} shows the results for the four SDSS bands.  
Because the mean surface brightness depends on waveband, we set 
$\log_{10}I = 0.4(27 - \mu_o + \langle\mu_o\rangle - \langle\mu_g\rangle)$
when making the plots, so as to facilitate comparison with 
Bender et al. (1992) and Burstein et al. (1997).  
The dashed line in the upper right corner of each panel shows 
$\kappa_1+\kappa_2=8$, what Burstein et al. termed the `zone of 
avoidance'.  Had we not accounted for the fact that the mean surface 
brightness is different in the different bands, then the galaxies 
would populate this zone.  

The magnitude limit is clearly visible in the lower right corner of 
the $\kappa_3$--$\kappa_1$ projection; we have not made any correction 
for it.  
When the sample is split by color, the redder galaxies appear to 
follow a tighter relation than the bluer galaxies, and they also tend 
to lie slightly closer to the zone of avoidance.  We leave quantifying 
and interpreting these trends to future work.

\section{Discussion and conclusions}\label{discuss}
We have studied the Fundamental Plane populated by $\sim 9000$ 
early-type galaxies over the redshift range $0\le z\le 0.3$ in 
the $g^*$, $r^*$, $i^*$ and $z^*$ bands.  If this Fundamental Plane 
is defined by minimizing the residuals orthogonal to it, then 
$R_o\propto \sigma^{1.5}I_o^{-0.77}$ (see Table~\ref{fpcoeffs} for 
the exact coefficients).  The Fundamental Plane is remarkably similar 
in the different bands (Figure~\ref{fig:FPgriz}), and appears to be 
slightly warped in the shorter wavebands (Figure~\ref{fig:warped}).  
Residuals with respect to the direct fit (i.e., the FP defined by 
minimizing the residuals in the direction of $\log_{10}R_o$) do not 
correlate with either velocity dispersion or color, whereas residuals 
from the orthogonal fit correlate with both (Figures~\ref{fig:FPresidR}).  
This correlation with $\sigma$ is simply a projection effect (see 
Figure~\ref{fig:FPsigR} and related discussion), whereas the correlation 
with color is mainly due to the fact that color and $\sigma$ are strongly 
correlated (Paper~IV). 
The Fundamental Plane is intrinsically slightly thinner in 
the redder wavebands.  This thickness is sometimes expressed in 
terms of the accuracy to which the FP can provide redshift-independent 
distance estimates---this is about 20\%.  If the thickness is expressed 
as a scatter in the mass-to-light ratio at fixed size and velocity 
dispersion, then this scatter is about 30\%.  

The simplest virial theorem prediction for the shape of the 
Fundamental Plane is that $R_o\propto \sigma^2/I_o$.  
This assumes that the observed velocity dispersion $\sigma^2$ is 
proportional to the kinetic energy $\sigma_{vir}^2$ which enters the 
virial theorem.  Busarello et al. (1997) argue that in their data 
$\log_{10}\sigma = (1.28\pm 0.11)\log_{10}\sigma_{vir} - 0.58$, so 
that $\sigma^{1.5}\propto\sigma_{vir}^{1.92}$.  Since the coefficient 
of $\sigma$ in the Fundamental Plane we find in all four bands is 
$\sim 1.5$, it would be interesting to see if the kinetic energy for the 
galaxies in our sample scales as it did in Busarello et al.'s sample.  
To do this, measurements of the velocity dispersion profiles of 
(a subsample of) the galaxies in our sample are required.  

Tests for passive luminosity evolution which use the Fundamental 
Plane are severly affected by selection effects and the choice of the 
fiducial Fundamental Plane against which to measure the evolution 
(Figures~\ref{fig:FPzG}--\ref{fig:FPresidmuz}).  
These tests suggest that the surface brightnesses of galaxies at 
higher redshifts in our sample are brighter than those of similar 
galaxies nearby.  The amount of brightening is consistent with the 
luminosity evolution estimated in Paper~II.  

The way in which galaxies scatter from the Fundamental Plane correlates 
weakly with their local environment (Figure~\ref{FPenviron}).  If this 
is caused entirely by differences in surface brightness, then galaxies 
in overdense regions are slightly fainter.  If so, then single-age 
stellar population models suggest that early-type galaxies in denser 
regions formed at higher redshift.  However, it may be that, the 
velocity dispersions are higher in denser regions (Paper~II).  
A larger sample is necessary to make a more definitive statement.

By the time the Sloan Digital Sky Survey is complete, the uncertainty 
in the K-corrections, which prevent us at the present time from making 
more precise quantitative statements about the evolution of the luminosities 
and colors, will be better understood.  In addition, the size of the sample 
will have increased by more than an order of magnitude.  This will allow 
us to provide a more quantitative study of the effects of environment than 
we are able to at the present time.

\vspace{1cm}

We would like to thank S. Charlot for making his stellar population 
synthesis predictions for the SDSS filters available to the 
collaboration and N. Benitez for making his package available.  
We thank D. Kelson and M. Pahre for helpful discussions.  

Funding for the creation and distribution of the SDSS Archive has been 
provided by the Alfred P. Sloan Foundation, the Participating Institutions, 
the National Aeronautics and Space Administration, the National Science 
Foundation, the U.S. Department of Energy, the Japanese Monbukagakusho, 
and the Max Planck Society. The SDSS Web site is http://www.sdss.org/.

The SDSS is managed by the Astrophysical Research Consortium (ARC) for
the Participating Institutions. The Participating Institutions are The
University of Chicago, Fermilab, the Institute for Advanced Study, the
Japan Participation Group, The Johns Hopkins University, Los Alamos
National Laboratory, the Max-Planck-Institute for Astronomy (MPIA),
the Max-Planck-Institute for Astrophysics (MPA), New Mexico State
University, University of Pittsburgh, Princeton University, the United
States Naval Observatory, and the University of Washington.

{}

\end{document}